\documentclass[11pt]{article}
\usepackage{graphicx}
\usepackage{amsmath,amssymb,amsfonts,amsthm}
\usepackage{setspace}
\usepackage{subfig}
\usepackage{booktabs}
\usepackage{multirow}
\usepackage{hyperref}
\usepackage{float}
\usepackage{color}
\hypersetup{
    colorlinks,%
    citecolor=black,%
    filecolor=black,%
    linkcolor=black,%
    urlcolor=blue
}
\doublespace
\numberwithin{equation}{section}

\usepackage[top=1.25 in, bottom=1.25 in, left=1.25 in , right=1.25 in]{geometry}
\usepackage{titlesec}
\titleformat{\section}{\large\bfseries}{\thesection}{1em}{}

\usepackage{enumitem}
\setdescription{leftmargin=\parindent,labelindent=\parindent}

\newtheorem{theorem}{Theorem}
\newtheorem{lemma}{Lemma}
\newtheorem{proposition}{Proposition}

\newtheorem{assumption}{Assumption}

\newenvironment{remark}[1][Remark]{\begin{trivlist}
\item[\hskip \labelsep {\bfseries #1}]}{\end{trivlist}}

\DeclareMathOperator*{\argmax}{arg\,max}
\DeclareMathOperator*{\argmin}{arg\,min}

\makeatletter
\renewcommand\@biblabel[1]{}
\makeatother

\begin{document}

\title{\Large{Asymptotic Refinements of a Misspecification-Robust Bootstrap for Generalized Empirical Likelihood Estimators}}
\author{Seojeong Lee
	\footnote{School of Economics, UNSW Business School, University of New South Wales, Sydney NSW 2052 Australia, Tel. (+61) 2 9385 3325, Fax. (+61) 2 9313 6337, Email \href{mailto:jay.lee@unsw.edu.au}{jay.lee@unsw.edu.au}, Web \href{https://sites.google.com/site/misspecifiedjay/}{https://sites.google.com/site/misspecifiedjay/}}\\
	\textit{\normalsize{University of New South Wales}}}

\date{\normalsize{Accepted for publication at the \textit{Journal of Econometrics}}}

\maketitle

\vspace{-1em}

\begin{abstract}
I propose a nonparametric iid bootstrap procedure for the empirical likelihood, the exponential tilting, and the exponentially tilted empirical likelihood estimators that achieves asymptotic refinements for $t$ tests and confidence intervals, and Wald tests and confidence regions based on such estimators. Furthermore, the proposed bootstrap is robust to model misspecification, i.e., it achieves asymptotic refinements regardless of whether the assumed moment condition model is correctly specified or not. This result is new, because asymptotic refinements of the bootstrap based on these estimators have not been established in the literature even under correct model specification. Monte Carlo experiments are conducted in dynamic panel data setting to support the theoretical finding. As an application, bootstrap confidence intervals for the returns to schooling of Hellerstein and Imbens (1999) are calculated. The result suggests that the returns to schooling may be higher. \\

\end{abstract}

\section{Introduction}
\label{S_Intro}

This paper establishes asymptotic refinements of the nonparametric iid bootstrap for $t$ tests and confidence intervals (CI's), and Wald tests and confidence regions based on the empirical likelihood (EL), the exponential tilting (ET), and the exponentially tilted empirical likelihood (ETEL) estimators. This is done without recentering the moment function in implementing the bootstrap, which has been considered as a critical procedure for overidentified moment condition models. Moreover, the proposed bootstrap is robust to misspecification, i.e., the resulting bootstrap tests and CI's achieve asymptotic refinements for the true parameter when the model is correctly specified, and the same rate of refinements is achieved for the pseudo-true parameter when misspecified. This is a new result because in the literature, there is no formal proof for asymptotic refinements of the bootstrap for EL, ET, or ETEL estimators even under correct specification. In fact, any bootstrap procedure with recentering for these estimators would be inconsistent if the model is misspecified because recentering imposes the correct model specification in the sample. This paper is motivated by three questions: (i) Why these estimators? (ii) Why bootstrap? (iii) Why care about misspecification?

First of all, EL, ET, and ETEL estimators are used to estimate a finite dimensional parameter characterized by a moment condition model. Traditionally, the generalized method of moments (GMM) estimators of Hansen (1982) have been used to estimate such models. However, it is well known that the two-step GMM may suffer from finite sample bias and inaccurate first-order asymptotic approximation to the finite sample distribution of the estimator when there are many moments, the model is non-linear, or instruments are weak. See Altonji and Segal (1996) and Hansen, Heaton, and Yaron (1996) among others on this matter.

Generalized empirical likelihood (GEL) estimators of Newey and Smith (2004) are alternatives to GMM as they have smaller asymptotic bias. GEL circumvents the estimation of the optimal weight matrix, which has been considered as a significant source of poor finite sample performance of the two-step GMM. GEL includes the EL estimator of Owen (1988, 1990), Qin and Lawless (1994), and Imbens (1997), the ET estimator of Kitamura and Stutzer (1997) and Imbens, Spady, and Johnson (1998), the continuously updating (CU) estimator of Hansen, Heaton, and Yaron (1996), and the minimum Hellinger distance estimator (MHDE) of Kitamura, Otsu, and Evdokimov (2013). Newey and Smith (2004) show that EL has the most favorable higher-order asymptotic properties than other GEL estimators. Although EL is preferable to other GEL estimators as well as GMM, its nice properties no longer hold under misspecification. In contrast, ET is often considered as robust to misspecification. Schennach (2007) proposes the ETEL estimator that shares the same higher-order property with EL under correct specification while possessing robustness of ET under misspecification. Hence, this paper considers the most widely used, EL, the most robust, ET, and a hybrid of the two, ETEL.\footnote{Precisely speaking, ETEL is not a GEL estimator. However, the analysis is quite similar because it is a combination of two GEL estimators. Therefore, this paper uses the term ``GEL'' to include ETEL as well as EL and ET to save space and to prevent any confusion.} An extension of the result to other GEL estimators is possible, but not attempted to make the argument succinct.

Secondly, many efforts have been made to accurately approximate the finite sample distribution of GMM. These include analytic correction of the GMM standard errors by Windmeijer (2005) and the bootstrap by Hahn (1996), Hall and Horowitz (1996), Andrews (2002), Brown and Newey (2002), Inoue and Shintani (2006), Allen, Gregory, and Shimotsu (2011), Lee (2014), among others. The bootstrap tests and CI's based on the GMM estimators achieve asymptotic refinements over the first-order asymptotic tests and CI's, which means their actual test rejection probability and CI coverage probability have smaller errors than the asymptotic tests and CI's. In particular, Lee (2014) applies a similar idea of non-recentering to GMM by using Hall and Inoue (2003)'s misspecification-robust variance estimators to achieve the same sharp rate of refinements with Andrews (2002).

Although GEL estimators are favorable alternatives to GMM, there is little evidence that the finite sample distribution of GEL test statistics is well approximated by the first-order asymptotics. Guggenberger and Hahn (2005) and Guggenberger (2008) find by simulation studies that the first-order asymptotic approximation to the finite sample distribution of EL estimators may be poor. Thus, it is natural to consider bootstrap $t$ tests and CI's based on GEL estimators to improve upon the first-order asymptotic approximation. However, few published papers deal with bootstrapping for GEL. Brown and Newey (2002) and Allen, Gregory, and Shimotsu (2011) employ the EL implied probability in resampling for GMM, but not for GEL. Canay (2010) shows the validity of a bootstrap procedure for the EL ratio statistic in the moment inequality setting. Kundhi and Rilstone (2012) argue that analytical corrections by Edgeworth expansion of the distribution of GEL estimators work well compared to the bootstrap, but they assume correct model specification.

Lastly, the validity of inferences and CI's critically depends on the correctly specified model assumption. Although model misspecification can be asymptotically detected by an overidentifying restrictions test, there is always a possibility that one does not reject a misspecified model or reject a correctly specified model in finite sample. Moreover, there is a view that all models are misspecified and will be rejected asymptotically. The consequences of model misspecification are twofold: a potentially biased probability limit of the estimator and a different asymptotic variance. The former is called the pseudo-true value, and it is impossible to correct the bias in general. Nevertheless, there are cases such that the pseudo-true values are still the object of interest: see Hansen and Jagannathan (1997), Hellerstein and Imbens (1999), Bravo (2010), and Almeida and Garcia (2012). GEL pseudo-true values are less arbitrary than GMM ones because the latter depend on a weight matrix, which is an arbitrary choice by a researcher. In contrast, each of the GEL pseudo-true values can be interpreted as a unique minimizer of a well-defined discrepancy measure, e.g. Schennach (2007).

The asymptotic variance of the estimator, however, can be consistently estimated even under misspecification. If a researcher wants to minimize the consequence of model misspecification, a misspecification-robust variance estimator should be used for $t$ tests or CI's, and for Wald tests and confidence regions. The proposed bootstrap uses the misspecification-robust variance estimator for EL, ET, and ETEL in constructing the $t$ or Wald statistics. This makes the proposed bootstrap robust to misspecification without recentering, and enables researchers to make valid inferences and CI's against unknown misspecification.

The remainder of the paper is organized as follows. Section \ref{S_outline} explains the idea of non-recentering by using a misspecification-robust variance estimator. Section \ref{S_est} defines the estimators and test statistics. Section \ref{S_MRB} describes the nonparametric iid misspecification-robust bootstrap procedure. Section \ref{S_Main} states the assumptions and establishes asymptotic refinements of the misspecification-robust bootstrap. Section \ref{S_MC} presents Monte Carlo experiments. An application to estimate the returns to schooling of Hellerstein and Imbens (1999) is presented in Section \ref{S_HI}. Section \ref{S_Con} concludes the paper. Lemmas and proofs are collected in Appendix A. A longer version of Lemmas and Proofs is available at the author's website.

\section{Asymptotic Refinement without Recentering}
\label{S_outline}

How does the proposed procedure achieve asymptotic refinements without recentering? The key idea is to construct an asymptotically pivotal statistic regardless of misspecification. Bootstrapping an asymptotically pivotal statistic is critical to get asymptotic refinements of the bootstrap (Beran, 1988; Hall, 1992; Horowitz, 2001). 

Suppose that $\chi_{n}=\{X_{i}:i\leq n\}$ is an iid sample. Let $F$ be the corresponding cumulative distribution function (cdf). Let $\theta$ be a parameter of interest and $g(X_{i},\theta)$ be a moment function. The moment condition model is correctly specified if $H_{C}: Eg(X_{i},\theta_{0}) = 0$ holds for a unique $\theta_{0}$.\footnote{This definition is from Hall and Inoue (2003) and assumes point identification.} The hypothesis is denoted by $H_{C}$. The hypothesis of interest is $H_{0}:\theta = \theta_{0}.$ The usual $t$ statistic $T_{C}$ is asymptotically standard normal under $H_{0}$ \textit{and} $H_{C}$. 

Now define the bootstrap sample. Let $\chi_{n_{b}}^{*}=\{X_{i}^{*}:i\leq n_{b}\}$ be a random draw with replacement from $\chi_{n}$ according to the empirical distribution function (edf) $F_{n}$. In this section, I distinguish the sample size $n$ and the bootstrap sample size $n_{b}$, following Bickel and Freedman (1981). The bootstrap versions of $H_{C}$ and $H_{0}$ are $H_{C}^{*}: E^{*}g(X_{i}^{*},\hat{\theta})=0$ and $H_{0}^{*}: \theta=\hat{\theta}$, where $E^{*}$ is the expectation taken over the bootstrap sample and $\hat{\theta}$ is a GEL estimator. Note that $\hat{\theta}$ is considered as the true value in the bootstrap world. The bootstrap version of the usual $t$ statistic $T_{C}^{*}$, however, is not asymptotically pivotal conditional on the sample because $H_{C}^{*}$ is not satisfied in the sample if the model is overidentified: $E^{*}g(X_{i}^{*},\hat{\theta}) = n^{-1}\sum_{i=1}^{n}g(X_{i},\hat{\theta}) \neq 0.$ Thus, Hall and Horowitz (1996), Andrews (2002), and Brown and Newey (2002) recenter the bootstrap version of the moment function to satisfy $H_{C}^{*}$. The resulting $t$ statistic based on the recentered moment function, $T_{C,R}^{*}$, tends to the standard normal distribution as $n_{b}$ grows conditional on the sample almost surely, and asymptotic refinements of the bootstrap are achieved.

This paper takes a different approach. Instead of jointly testing $H_{C}$ and $H_{0}$, I solely focus on $H_{0}$, leaving that $H_{C}$ may not hold. If the model is misspecified, then there is no such $\theta$ that satisfies $H_{C}$, i.e. $Eg(X_{i},\theta) \neq 0, \forall \theta\in\Theta,$ where $\Theta$ is a compact parameter space. This may happen if the model is overidentified. Since there is no true value, the pseudo-true value $\theta_{0}$ should be defined. Instead of $H_{C}$, $\theta_{0}$ is defined as a unique minimizer of the population version of the empirical discrepancy used in the estimation. For EL, this discrepancy is the Kullback-Leibler Information Criterion (KLIC). For ET, it maximizes a quantity named entropy. This definition is more flexible since it includes correct specification as a special case when $H_{C}$ holds at $\theta_{0}$. Without assuming $H_{C}$, we can find regularity conditions for $\sqrt{n}-$consistency and asymptotic normality of $\hat{\theta}$ for the pseudo-true value $\theta_{0}$. Under misspecification and suitable regularity conditions, as the sample size grows, 
\begin{equation}
\sqrt{n}(\hat{\theta}-\theta_{0})\rightarrow_{d}N(0,\Sigma_{MR}).
\end{equation}
The asymptotic variance matrix $\Sigma_{MR}$ is different from the standard asymptotic variance matrix, but it coincides with the standard one under correct specification. $\Sigma_{MR}$ can be consistently estimated using the formula given in the next section. Let $\hat{\Sigma}_{MR}$ be a consistent estimator for $\Sigma_{MR}$. The misspecification-robust $t$ statistic $T_{MR}$ is studentized with $\hat{\Sigma}_{MR}$. Thus, $T_{MR}$ is asymptotically standard normal under $H_{0}$, without assuming $H_{C}$.

Similarly, we construct the bootstrap version of the $t$ statistic using the same formula as the sample misspecification-robust $t$ statistic. Conditional on the sample almost surely, $T_{MR}^{*}$ tends to the standard normal distribution as $n_{b}$ grows under $H_{0}^{*}$. Since the conditional asymptotic distribution does not depend on $H_{C}^{*}$, we need not recenter the bootstrap moment function to satisfy $H_{C}^{*}$. In other words, the misspecification-robust $t$ statistic $T_{MR}$ is asymptotically pivotal under $H_{0}$, while the usual $t$ statistic $T_{C}$ is asymptotically pivotal under $H_{0}$ \textit{and} $H_{C}$. This paper develops a theory for bootstrapping $T_{MR}$, instead of $T_{C}$. Note that both can be used to test the null hypothesis $H_{0}:\theta=\theta_{0}$ under correct specification. Under misspecification, however, only $T_{MR}$ can be used to test $H_{0}$ because $T_{C}$ is not asymptotically pivotal. This is useful when the pseudo-true value is an interesting object even if the model is misspecified.

To find the formula for $\Sigma_{MR}$, I use a just-identified system of the first-order conditions (FOC's) of EL, ET, and ETEL estimators. This idea is not new, though. Schennach (2007) uses the same idea to find the asymptotic variance matrix of the ETEL estimator robust to misspecification. For GMM estimators, the idea of rewriting the overidentified GMM as a just-identified system appears in Imbens (1997,2002) and Chamberlain and Imbens (2003).

A natural question is whether we can use GEL implied probabilities to construct the cdf estimator $\hat{F}$ and use it instead of the edf $F_{n}$ in resampling. This is possible when the moment condition is correctly specified. For instance, Brown and Newey (2002) argue that using the EL-estimated cdf $\hat{F}_{EL}(z)\equiv\sum_{i}\mathbf{1}(X_{i}\leq z)p_{i}$, where $p_{i}$ is the EL implied probability, in place of the edf $F_{n}$ in resampling would improve efficiency of bootstrapping for GMM. Their argument relies on the fact that $\hat{F}_{EL}$ is an efficient estimator of the true cdf $F$. If the moment condition is misspecified, however, then the cdf estimator based on the implied probability is inconsistent for $F$ because $E_{\hat{F}}g(X_{i},\hat{\theta})=\sum_{i}p_{i}g(X_{i},\hat{\theta})=0$ holds even in large sample, while $Eg(X_{i},\theta_{0})\neq 0$.\footnote{Bootstrapping the EL ratio test statistics of Owen (1988, 1990) under misspecification may not be a good idea for this reason, because the EL likelihood function is a product of EL implied probabilities that are inconsistent.} In contrast, the edf $F_{n}$ is uniformly consistent for $F$ regardless of whether the moment condition holds or not by Glivenko-Cantelli Theorem. For this reason, I mainly focus on resampling from $F_{n}$ rather than $\hat{F}$ in this paper.\footnote{However, a shrinkage-type cdf estimator combining $F_{n}$ and $\hat{F}$, similar to Antoine, Bonnal, and Renault (2007), can be used to improve both robustness and efficiency. For example, a shrinkage that has the form
$\pi_{i} = \epsilon_{n}\cdot p_{i} + (1-\epsilon_{n})\cdot n^{-1}$, where $\epsilon_{n}\rightarrow 0$ as $n$ grows, would work with the proposed misspecification-robust bootstrap because $E_{\pi}g(X_{i},\hat{\theta}) = (1-\epsilon_{n})n^{-1}\sum_{i}g(X_{i},\hat{\theta}) \neq 0,$ where the expectation is taken with respect to $\hat{F}_{\pi}(z)\equiv\sum_{i}\mathbf{1}(X_{i}\leq z)\pi_{i}$.} 

\section{Estimators and Test Statistics}
\label{S_est}

Let $g(X_{i},\theta)$ be an $L_{g}\times 1 $ moment function where $\theta\in\Theta\subset \mathbf{R}^{L_{\theta}}$ is a parameter of interest, where $L_{g}\geq L_{\theta}$. Let $G^{(j)}(X_{i},\theta)$ denote the vectors of partial derivatives with respect to $\theta$ of order $j$ of $g(X_{i},\theta)$. In particular, $G^{(1)}(X_{i},\theta)\equiv G(X_{i},\theta)\equiv(\partial/\partial\theta')g(X_{i},\theta)$ is an $L_{g}\times L_{\theta}$ matrix and $G^{(2)}(X_{i},\theta)\equiv(\partial/\partial\theta')vec\{G(X_{i},\theta)\}$ is an $L_{g}L_{\theta}\times L_{\theta}$ matrix, where $vec\{\cdot\}$ is the vectorization of a matrix. To simplify notation, write $g_{i}(\theta)=g(X_{i},\theta)$, $G_{i}^{(j)}(\theta) = G^{(j)}(X_{i},\theta)$, $\hat{g}_{i}=g(X_{i},\hat{\theta})$, and $\hat{G}_{i}^{(j)}=G^{(j)}(X_{i},\hat{\theta})$ for $j=1,...,d+1$, where $\hat{\theta}$ is EL, ET or ETEL estimator. In addition, let $g_{i0}=g_{i}(\theta_{0})$ and $G_{i0}=G_{i}(\theta_{0})$, where $\theta_{0}$ is the (pseudo-)true value.

\subsection{Empirical Likelihood and Exponential Tilting Estimators}

Following the notation of Newey and Smith (2004) and Anatolyev (2005), let $\rho(\nu)$ be a concave function in a scalar $\nu$ on the domain that contains zero. For EL, $\rho(\nu)=\log(1-\nu)$ for $\nu\in(-\infty,1)$. For ET, $\rho(\nu) = 1-e^{\nu}$ for $\nu\in\mathbf{R}$. In addition, let $\rho_{j}(\nu)=\partial^{j}\rho(\nu)/\partial\nu^{j}$ for $j=0,1,2,\cdots$. 

The EL or the ET estimator, $\hat{\theta}$, and the corresponding Lagrange multiplier, $\hat{\lambda}$, solve a saddle point problem
\begin{equation}
\min_{\theta\in\Theta}\max_{\lambda}n^{-1}\sum_{i=1}^{n}\rho(\lambda'g_{i}(\theta)).
\label{GEL}
\end{equation}
The FOC's for $(\hat{\theta},\hat{\lambda})$ are
\begin{equation} \underset{L_{\theta}\times 1}{0}=n^{-1}\sum_{i=1}^{n}\rho_{1}(\hat{\lambda}'\hat{g}_{i})\hat{G}_{i}'\hat{\lambda},\hspace{1em}\underset{L_{g}\times 1}{0}=n^{-1}\sum_{i=1}^{n}\rho_{1}(\hat{\lambda}'\hat{g}_{i})\hat{g}_{i}.
\end{equation}
A useful by-product of the estimation is the implied probabilities. The EL and the ET implied probabilities for the observations are, for $i=1,...,n$,
\begin{equation}
\text{EL: } p_{i} = \frac{1}{n(1-\hat{\lambda}'\hat{g}_{i})},\hspace{1em}\text{ET: } p_{i} = \frac{e^{\hat{\lambda}'\hat{g}_{i}}}{\sum_{j=1}^{n}e^{\hat{\lambda}'\hat{g}_{j}}}.
\end{equation}

The FOC's hold regardless of model misspecification and form a just-identified moment condition. Let $\psi(X_{i},\beta)$ be a $(L_{\theta}+L_{g})\times 1$ vector such that
\begin{equation}
\psi(X_{i},\beta)\equiv\left[
                      \begin{array}{c}
                        \psi_{1}(X_{i},\beta) \\
                        \psi_{2}(X_{i},\beta) \\
                      \end{array}
                    \right]
                   =\left[
                    \begin{array}{c}
                    \rho_{1}(\lambda'g_{i}(\theta))G_{i}(\theta)'\lambda \\
                    \rho_{1}(\lambda'g_{i}(\theta))g_{i}(\theta) \\
                    \end{array}
                  \right].
\label{GEL_EE}
\end{equation}
Then, the EL or the ET estimator and the corresponding Lagrange multiplier denoted by an augmented vector, $\hat{\beta}=(\hat{\theta}',\hat{\lambda}')'$, solves $n^{-1}\sum_{i=1}^{n}\psi(X_{i},\hat{\beta})=0$. In the limit, the pseudo-true value $\beta_{0}=(\theta_{0}',\lambda_{0}')'$ solves the population version of the FOC's:
\begin{equation} \underset{L_{\theta}\times 1}{0}=E\rho_{1}(\lambda_{0}'g_{i0})G_{i0}'\lambda_{0},\hspace{1em}\underset{L_{g}\times 1}{0}=E\rho_{1}(\lambda_{0}'g_{i0})g_{i0}.
\end{equation}
The asymptotic distribution of $\hat{\beta}=(\hat{\theta}',\hat{\lambda}')'$ can be derived by using standard asymptotic theory of just-identified GMM, e.g. Newey and McFadden (1994).

For EL, Chen, Hong, and Shum (2007) provide regularity conditions for $\sqrt{n}$-consistency and asymptotic normality under misspecification. In particular, they assume that the moment function is uniformly bounded:
\begin{equation}
\textbf{UBC: } \sup_{\theta\in\Theta,x\in\chi}\|g(x,\theta)\|<\infty \hspace{0.5em}\text{ and } \inf_{\theta\in\Theta,\lambda\in\Lambda(\theta),x\in\chi}(1-\lambda'g(x,\theta))>0,
\label{UBC}
\end{equation}
where $\Theta$ and $\Lambda(\theta)$ are compact sets and $\chi$ is the support of $X_{1}$. Schennach (2007) shows that the EL estimator is no longer $\sqrt{n}$-consistent if the moment function is unbounded for any $\theta$. Nevertheless, if the data is bounded or the moment function is constructed to satisfy UBC, then the EL estimator would be $\sqrt{n}$-consistent for the pseudo-true value and the bootstrap can be implemented. For ET, UBC is not required.

Assuming regularity conditions such as Assumption 3 of Chen, Hong, and Shum (2007) for EL, and Assumption 3 of Schennach (2007) for ET.\footnote{Schennach's assumptions are for ETEL but can be easily modified for ET. First, Assumption 3(2) needs to be replaced with the ET saddle-point problem. In addition, we only require $k_{2}=0,1,2$ instead of $k_{2}=0,1,2,3,4$ in Assumption 3(6).}, we have the following proposition:
\begin{proposition}
Let $\hat{\beta}=(\hat{\theta}',\hat{\lambda}')'$ be either the EL or the ET estimator and its Lagrange multiplier, and $\beta_{0}=(\theta_{0}',\lambda_{0}')'$ be the corresponding pseudo-true value. Then,
\[\sqrt{n}(\hat{\beta}-\beta_{0})\rightarrow_{d}N(0,\Gamma^{-1}\Psi(\Gamma')^{-1}),\]
where $\Gamma = E(\partial/\partial\beta')\psi(X_{i},\beta_{0})$ and $\Psi=E\psi(X_{i},\beta_{0})\psi(X_{i},\beta_{0})'$.
\label{PropEL}
\end{proposition}

The Jacobian matrix for EL or ET is given by
\begin{equation}
\frac{\partial\psi(X_{i},\beta)}{\partial\beta'}=\left[
                                                       \begin{array}{cc}
                                                       (\partial/\partial\theta')\psi_{1}(X_{i},\beta) & (\partial/\partial\lambda')\psi_{1}(X_{i},\beta) \\
                                                       (\partial/\partial\theta')\psi_{2}(X_{i},\beta) & (\partial/\partial\lambda')\psi_{2}(X_{i},\beta) \\
                                                       \end{array}
                                                       \right],
\end{equation}
where
\begin{eqnarray}
\frac{\partial\psi_{1}(X_{i},\beta)}{\partial\theta'} &=& \rho_{1}(\lambda'g_{i}(\theta))(\lambda'\otimes I_{L_{\theta}})G_{i}^{(2)}(\theta) + \rho_{2}(\lambda'g_{i}(\theta))G_{i}(\theta)'\lambda\lambda'G_{i}(\theta),\\
\nonumber \frac{\partial\psi_{1}(X_{i},\beta)}{\partial\lambda'} &=& \frac{\partial\psi_{2}(X_{i},\beta)}{\partial\theta} = \rho_{1}(\lambda'g_{i}(\theta))G_{i}(\theta)' + \rho_{2}(\lambda'g_{i}(\theta))G_{i}(\theta)'\lambda g_{i}(\theta)',\\
\nonumber \frac{\partial\psi_{2}(X_{i},\beta)}{\partial\lambda'} &=&  \rho_{2}(\lambda'g_{i}(\theta))g_{i}(\theta)g_{i}(\theta)'.
\end{eqnarray}
$\Gamma$ and $\Psi$ can be estimated by
\begin{equation}
\hat{\Gamma}=n^{-1}\sum_{i=1}^{n}\frac{\partial\psi(X_{i},\hat{\beta})}{\partial\beta'}\hspace{0.5em}\text{ and }\hspace{0.5em}\hat{\Psi}=n^{-1}\sum_{i=1}^{n}\psi(X_{i},\hat{\beta})\psi(X_{i},\hat{\beta})',
\label{VAREST}
\end{equation}
respectively. The upper left $L_{\theta}\times L_{\theta}$ submatrix of $\Gamma^{-1}\Psi(\Gamma')^{-1}$, denoted by $\Sigma_{MR}$, is the asymptotic variance matrix of $\sqrt{n}(\hat{\theta}-\theta_{0})$. This matrix coincides with the usual asymptotic variance matrix $\Sigma_{C}=(EG_{i0}'(Eg_{i0}g_{i0}')^{-1}EG_{i0})^{-1}$ under correct specification. Let $\hat{\Sigma}_{MR}$ be the corresponding submatrix of the variance estimator $\hat{\Gamma}^{-1}\hat{\Psi}(\hat{\Gamma}')^{-1}$. Even under correct specification, $\hat{\Sigma}_{MR}$ is different from $\hat{\Sigma}_{C}$, the conventional variance estimator, because $\hat{\Sigma}_{MR}$ contains additional terms which are assumed away in $\hat{\Sigma}_{C}$.
 
\subsection{Exponentially Tilted Empirical Likelihood Estimator}

Schennach (2007) proposes the ETEL estimator which is robust to misspecification without UBC, while it maintains the same nice higher-order properties with EL under correct specification. The ETEL estimator and the Lagrange multiplier $(\hat{\theta},\hat{\lambda})$ solve
\begin{equation}
\argmin_{\theta\in\Theta}-n^{-1}\sum_{i=1}^{n}\log n\hat{w}_{i}(\theta),\hspace{1em}\hat{w}_{i}(\theta)=\frac{e^{\hat{\lambda}(\theta)'g_{i}(\theta)}}{\sum_{j=1}^{n}e^{\hat{\lambda}(\theta)'g_{j}(\theta)}},
\label{ETEL}
\end{equation}
where $\hat{\lambda}\equiv \hat{\lambda}(\hat{\theta})$ and
\begin{equation}
\hat{\lambda}(\theta)=\argmax_{\lambda}-n^{-1}\sum_{i=1}^{n}e^{\lambda'g_{i}(\theta)}.
\end{equation}
This estimator is a hybrid of the EL estimator and the ET implied probability. Equivalently, the ETEL estimator $\hat{\theta}$ minimizes the objective function
\begin{equation}
\hat{l}_{n}(\theta) = \log \left(n^{-1}\sum_{i=1}^{n}e^{\hat{\lambda}(\theta)'(g_{i}(\theta)-\bar{g}_{n}(\theta))}\right),
\label{ETEL1}
\end{equation}
where $\bar{g}_{n}(\theta)=n^{-1}\sum_{i=1}^{n}g_{i}(\theta)$. To derive the asymptotic distribution of the ETEL estimator, Schennach introduces auxiliary parameters to formulate the problem into a just-identified GMM. Let $\beta = (\theta',\lambda',\kappa',\tau)'$, where $\kappa\in\mathbf{R}^{L_{g}}$ and $\tau\in\mathbf{R}$. By Lemma 9 of Schennach (2007), the ETEL estimator $\hat{\theta}$ is given by the subvector of $\hat{\beta}=(\hat{\theta}',\hat{\lambda}',\hat{\kappa}',\hat{\tau})'$, that solves
\begin{eqnarray}
n^{-1}\sum_{i=1}^{n}\psi(X_{i},\hat{\beta})=0,
\label{OBJ}
\end{eqnarray}
where
\begin{equation}
\psi(X_{i},\beta)\equiv\left[
                      \begin{array}{c}
                        \psi_{1}(X_{i},\beta) \\
                        \psi_{2}(X_{i},\beta) \\
                        \psi_{3}(X_{i},\beta) \\
                        \psi_{4}(X_{i},\beta) \\
                      \end{array}
                    \right]
                   =\left[
                    \begin{array}{c}
                      e^{\lambda'g_{i}(\theta)}G_{i}(\theta)'\left(\kappa + \lambda g_{i}(\theta)'\kappa - \lambda\right) + \tau G_{i}(\theta)'\lambda \\
                      (\tau-e^{\lambda'g_{i}(\theta)})\cdot g_{i}(\theta) + e^{\lambda'g_{i}(\theta)}\cdot g_{i}(\theta)g_{i}(\theta)'\kappa \\
                      e^{\lambda'g_{i}(\theta)}\cdot g_{i}(\theta) \\
                      e^{\lambda'g_{i}(\theta)}-\tau \\
                    \end{array}
                  \right].
\label{ETEL_EE}
\end{equation}
Note that the estimators of the auxiliary parameters, $\hat{\kappa}$ and $\hat{\tau}$ are given by
\begin{equation}
\hat{\tau} = n^{-1}\sum_{i=1}^{n}e^{\hat{\lambda}'\hat{g}_{i}}\hspace{0.5em}\text{and}\hspace{0.5em}\hat{\kappa} = -\left(n^{-1}\sum_{i=1}^{n}\frac{e^{\hat{\lambda}'\hat{g}_{i}}}{\hat{\tau}}\hat{g}_{i}\hat{g}_{i}'\right)^{-1}\hat{\bar{g}}_{n},
\end{equation}
where $\hat{\bar{g}}_{n}=n^{-1}\sum_{i=1}^{n}\hat{g}_{i}$. The probability limit of $\hat{\beta}$ is the pseudo-true value $\beta_{0}= (\theta_{0}',\lambda_{0}',\kappa_{0}',\tau_{0})'$ that solves $E\psi(X_{i},\beta_{0})=0$. In particular, a function $\lambda_{0}(\theta)$ is the solution to $Ee^{\lambda'g_{i}(\theta)}g_{i}(\theta)=0$, where $\lambda_{0}\equiv\lambda_{0}(\theta_{0})$ and $\theta_{0}$ is a unique minimizer of the population objective function:
\begin{equation}
l_{0}(\theta) = \log \left(Ee^{\lambda_{0}(\theta)'(g_{i}(\theta)-Eg_{i}(\theta))}\right).
\end{equation}
By Theorem 10 of Schennach,
\begin{equation}
\sqrt{n}(\hat{\beta}-\beta_{0})\rightarrow_{d}N(0,\Gamma^{-1}\Psi(\Gamma')^{-1}),
\end{equation}
where $\Gamma = E(\partial/\partial\beta')\psi(X_{i},\beta_{0})$ and $\Psi=E\psi(X_{i},\beta_{0})\psi(X_{i},\beta_{0})'$. 

$\Gamma$ and $\Psi$ are estimated by the same formula with \eqref{VAREST}. In order to estimate $\Gamma$, we need a formula of $(\partial/\partial\beta')\psi(X_{i},\beta)$. The partial derivative of $\psi_{1}(X_{i},\beta)$ is given by
\begin{equation}
\frac{\partial\psi_{1}(X_{i},\beta)}{\partial\beta'}=\left(
                                                         \begin{array}{cccc}
                                                              \underset{L_{\theta}\times L_{\theta}}{\frac{\partial\psi_{1}(X_{i},\beta)}{\partial\theta'}}& \underset{L_{\theta}\times L_{g}}{\frac{\partial\psi_{1}(X_{i},\beta)}{\partial\lambda'}} &\underset{L_{\theta}\times L_{g}}{\frac{\partial\psi_{1}(X_{i},\beta)}{\partial\kappa'}}& \underset{L_{\theta}\times 1}{\frac{\partial\psi_{1}(X_{i},\beta)}{\partial\tau}}\\
                                                         \end{array}
                                                       \right),
\end{equation}
where
{
\allowdisplaybreaks
\begin{eqnarray}
\frac{\partial\psi_{1}(X_{i},\beta)}{\partial\theta'} &=& e^{\lambda'g_{i}(\theta)}\left\{G_{i}(\theta)'(\kappa\lambda'+\lambda\kappa'+\lambda g_{i}(\theta)'\kappa\lambda'-\lambda\lambda')G_{i}(\theta) \right.\\
\nonumber && \left. + ((\kappa'+\kappa'g_{i}(\theta)\lambda'-\lambda')\otimes I_{L_{\theta}})G_{i}^{(2)(\theta)}\right\} + \tau(\lambda'\otimes I_{L_{\theta}})G_{i}^{(2)}(\theta),\\
\frac{\partial\psi_{1}(X_{i},\beta)}{\partial\lambda'} &=& e^{\lambda'g_{i}(\theta)}G_{i}(\theta)'\left\{( \lambda g_{i}(\theta)'\kappa+\kappa-\lambda) g_{i}(\theta)' + (g_{i}(\theta)'\kappa-1)I_{L_{g}}\right\}\\
\nonumber && +\tau G_{i}(\theta)',\\
\frac{\partial\psi_{1}(X_{i},\beta)}{\partial\kappa'} &=& e^{\lambda'g_{i}(\theta)}G_{i}(\theta)'(I_{L_{g}}+\lambda g_{i}(\theta)'),\\
\frac{\partial\psi_{1}(X_{i},\beta)}{\partial\tau} &=& G_{i}(\theta)'\lambda.
\end{eqnarray}
}
The partial derivative of $\psi_{2}(X_{i},\beta)$ is given by
\begin{equation}
\frac{\partial\psi_{2}(X_{i},\beta)}{\partial\beta'}=\left(
                                                         \begin{array}{cccc}
                                                           \underset{L_{g}\times L_{\theta}}{\frac{\partial\psi_{2}(X_{i},\beta)}{\partial\theta'}} &\underset{L_{g}\times L_{g}}{\frac{\partial\psi_{2}(X_{i},\beta)}{\partial\lambda'}}  & \underset{L_{g}\times L_{g}}{e^{\lambda'g_{i}(\theta)}g_{i}(\theta)g_{i}(\theta)'} & \underset{L_{g}\times 1}{g_{i}(\theta)}  
                                                         \end{array}
                                                       \right),
\end{equation}
where
\begin{eqnarray}
\frac{\partial\psi_{2}(X_{i},\beta)}{\partial\theta'}&=&\frac{\partial\psi_{1}(X_{i},\beta)}{\partial\lambda},\\
\frac{\partial\psi_{2}(X_{i},\beta)}{\partial\lambda'}&=&e^{\lambda'g_{i}(\theta)}g_{i}(\theta)g_{i}(\theta)'(\kappa g_{i}(\theta)'-I_{L_{g}}).
\end{eqnarray}
The partial derivative of $\psi_{3}(X_{i},\beta)$ is given by
\begin{equation}
\frac{\partial\psi_{3}(X_{i},\beta)}{\partial\beta'}=\left(
                                                         \begin{array}{cccc}
                                                           \underset{L_{g}\times L_{\theta}}{\frac{\partial\psi_{1}(X_{i},\beta)}{\partial\kappa}} & \underset{L_{g}\times L_{g}}{e^{\lambda'g_{i}(\theta)}g_{i}(\theta)g_{i}(\theta)'} &\underset{L_{g}\times L_{g}}{\mathbf{0}} &\underset{L_{g}\times 1}{\mathbf{0}}   \\
                                                         \end{array}
                                                       \right),
\end{equation}
and the partial derivative of $\psi_{4}(X_{i},\beta)$ is given by
\begin{equation}
\frac{\partial\psi_{4}(X_{i},\beta)}{\partial\beta'}=\left(
                                                         \begin{array}{cccc}
                                                           \underset{1\times L_{\theta}}{e^{\lambda'g_{i}(\theta)}\lambda'G_{i}(\theta)} & \underset{1\times L_{g}}{e^{\lambda'g_{i}(\theta)}g_{i}(\theta)'} & \underset{1\times L_{g}}{\mathbf{0}} & \underset{1\times 1}{-1} \\
                                                         \end{array}
                                                       \right).
\end{equation}                                                       
The upper left $L_{\theta}\times L_{\theta}$ submatrix of $\Gamma^{-1}\Psi(\Gamma')^{-1}$, denoted by $\Sigma_{MR}$, is the asymptotic variance matrix of $\sqrt{n}(\hat{\theta}-\theta_{0})$. Let $\hat{\Sigma}_{MR}$ be the corresponding submatrix of the variance estimator $\hat{\Gamma}^{-1}\hat{\Psi}(\hat{\Gamma}')^{-1}$. Again, $\Sigma_{MR}$ is different from $\Sigma_{C}$ in general under misspecification, but they become identical under correct specification.\footnote{Under correct specification, the asymptotic variance matrix $\Sigma_{C}$ is the same for EL, ET, and ETEL, which is the asymptotic variance matrix of the two-step efficient GMM.}

\subsection{Test statistics}

Let $\hat{\theta}$ be either the EL, the ET, or the ETEL estimator and let $\hat{\Sigma}_{MR}$ be the corresponding variance matrix estimator. Let $\theta_{r}$, $\theta_{0,r}$, and $\hat{\theta}_{r}$ denote the $r$th elements of $\theta$, $\theta_{0}$, and $\hat{\theta}$ respectively. Let $\hat{\Sigma}_{MR,r}$ denote the $r$th diagonal element of $\hat{\Sigma}_{MR}$. The $t$ statistic for testing the null hypothesis $H_{0}:\theta_{r}=\theta_{0,r}$ is
\begin{equation}
T_{MR} = \frac{\hat{\theta}_{r}-\theta_{0,r}}{\sqrt{\hat{\Sigma}_{MR,r}/n}}.
\end{equation}
Since $T_{MR}$ is studentized with the misspecification-robust variance estimator $\hat{\Sigma}_{MR,r}$, it has an asymptotic $N(0,1)$ distribution under $H_{0}$, without assuming the correct model, $H_{C}$. This is the source of achieving asymptotic refinements without recentering regardless of misspecification. In contrast, the usual $t$ statistic $T_{C}$ is studentized with $\hat{\Sigma}_{C}$, a non-robust variance estimator. Hence, it is not asymptotically pivotal if the model is misspecified. Note that the only difference between $T_{MR}$ and $T_{C}$ is the variance estimator.

We also consider the Wald statistic for multivariate tests and confidence regions. Let $\eta(\theta)$ be an $\mathbf{R}^{L_{\eta}}$-valued function that is continuously differentiable at $\theta_{0}$. The Wald statistic for testing $H_{0}:\eta(\theta_{0})=0$ against $H_{1}:\eta(\theta_{0})\neq0$ is 
\begin{equation}
\label{wald}
\mathcal{W}_{MR}=n\cdot \eta(\hat{\theta})'\left(\frac{\partial}{\partial\theta'}\eta(\hat{\theta})\hat{\Sigma}_{MR}\left(\frac{\partial}{\partial\theta'}\eta(\hat{\theta})\right)'\right)^{-1}\eta(\hat{\theta}).
\end{equation}
This Wald statistic is different from the conventional one because $\hat{\Sigma}_{MR}$ is used. Thus, its asymptotic distribution is a chi-square distribution with $L_{\eta}$ degrees of freedom, denoted by $\chi^{2}_{L_{\eta}}$, regardless of misspecification. 

Both one-sided and two-sided $t$ tests with asymptotic significance level $\alpha$ and CI's with asymptotic confidence level $1-\alpha$ are considered. The asymptotic one-sided $t$ test of $H_{0}: \theta_{r}\leq\theta_{0,r}$ against $H_{1}:\theta_{r}>\theta_{0,r}$ rejects $H_{0}$ if $T_{MR}>z_{\alpha}$, where $z_{\alpha}$ is the $1-\alpha$ quantile of the standard normal distribution. This one-sided test corresponds to the lower endpoint one-sided CI, $[\hat{\theta}_{r}- z_{\alpha}\sqrt{\hat{\Sigma}_{MR,r}/n},\infty)$. The asymptotic two-sided $t$ test of $H_{0}: \theta_{r}=\theta_{0,r}$ against $H_{1}:\theta_{r}\neq\theta_{0,r}$ rejects $H_{0}$ if $|T_{MR}|>z_{\alpha/2}$. The two-sided asymptotic CI is $[\hat{\theta}_{r}\pm z_{\alpha/2}\sqrt{\hat{\Sigma}_{MR,r}/n}]$. The asymptotic Wald test of $H_{0}:\eta(\theta_{0})=0$ against $H_{1}:\eta(\theta_{0})\neq0$ rejects the null if $\mathcal{W}_{MR}>z_{\mathcal{W},\alpha}$, where $z_{\mathcal{W},\alpha}$ is the $1-\alpha$ quantile of a chi-square distribution with $L_{\eta}$ degrees of freedom. The Wald-based asymptotic confidence region for $\eta(\theta_{0})$ is $\{\eta\in\mathbf{R}^{L_{\eta}}:n\cdot (\eta(\hat{\theta})-\eta)'((\partial\eta(\hat{\theta})/\partial\theta')\hat{\Sigma}_{MR}(\partial\eta(\hat{\theta})/\partial\theta')')^{-1}(\eta(\hat{\theta})-\eta)\leq z_{\mathcal{W},\alpha}\}$. All the tests and CI's have the correct asymptotic significance and confidence levels regardless of misspecification because they are based on the misspecification-robust test statistics.

\section{The Misspecification-Robust Bootstrap Procedure}
\label{S_MRB}

The nonparametric iid bootstrap is implemented by resampling $X_{1}^{*},\cdots,X_{n}^{*}$ randomly with replacement from the sample $X_{1},\cdots,X_{n}$. Based on the bootstrap sample, $\chi_{n}^{*}=\{X_{i}^{*}:i\leq n\}$, the bootstrap GEL estimator $\hat{\theta}^{*}$ solves \eqref{GEL} for EL or ET, and \eqref{ETEL} for ETEL. The bootstrap version of the variance matrix estimator is $\hat{\Gamma}^{*-1}\hat{\Psi}^{*}(\hat{\Gamma}^{*'})^{-1}$ which can be calculated using the same formula with \eqref{VAREST} using the bootstrap sample instead of the original sample. Let $\hat{\Sigma}^{*}_{MR}$ be the upper left $L_{\theta}\times L_{\theta}$ submatrix of $\hat{\Gamma}^{*-1}\hat{\Psi}^{*}(\hat{\Gamma}^{*'})^{-1}$. I emphasize that no additional corrections such as recentering as in Hall and Horowitz (1996) and Andrews (2002) are required.

The misspecification-robust bootstrap $t$ and Wald statistics are
\begin{eqnarray}
T_{MR}^{*} &=& \frac{\hat{\theta}_{r}^{*}-\hat{\theta}_{r}}{\sqrt{\hat{\Sigma}^{*}_{MR,r}/n}},\\
\mathcal{W}_{MR}^{*} &=& n\cdot (\eta(\hat{\theta}^{*})-\eta(\hat{\theta}))'\left(\frac{\partial}{\partial\theta'}\eta(\hat{\theta}^{*})\hat{\Sigma}_{MR}^{*}\left(\frac{\partial}{\partial\theta'}\eta(\hat{\theta}^{*})\right)'\right)^{-1}(\eta(\hat{\theta}^{*})-\eta(\hat{\theta})).
\end{eqnarray}
Let $z^{*}_{T,\alpha}$, $z^{*}_{|T|,\alpha}$, $z^{*}_{\mathcal{W},\alpha}$ denote the $1-\alpha$ quantile of $T_{MR}^{*}$, $|T_{MR}^{*}|$, and $\mathcal{W}^{*}_{MR}$, respectively. Let $P^{*}$ be the probability distribution of the bootstrap sample conditional on the sample. Following Andrews (2002), we define $z^{*}_{|T|,\alpha}$ to be a value that minimizes $|P^{*}(|T_{MR}^{*}|\leq z)-(1-\alpha)|$ over $z\in \mathbf{R}$, because the distribution of $|T_{MR}^{*}|$ is discrete. The definitions of $z^{*}_{T,\alpha}$ and $z^{*}_{\mathcal{W},\alpha}$ are analogous. Each of the following bootstrap tests are of asymptotic significance level $\alpha$. The one-sided bootstrap $t$ test of $H_{0}: \theta_{r}\leq\theta_{0,r}$ against $H_{1}:\theta_{r}>\theta_{0,r}$ rejects $H_{0}$ if $T_{MR}>z_{T,\alpha}^{*}$. The symmetric two-sided bootstrap $t$ test of $H_{0}:\theta_{r}=\theta_{0,r}$ versus $H_{1}:\theta_{r}\neq \theta_{0,r}$ rejects if $|T_{MR}|>z^{*}_{|T|,\alpha}$. The equal-tailed two-sided bootstrap $t$ test of the same hypotheses rejects if $T_{MR}<z^{*}_{T,1-\alpha/2}$ or $T_{MR}>z^{*}_{T,\alpha/2}$. The bootstrap Wald test of $H_{0}:\eta(\theta_{0})=0$ against $H_{1}:\eta(\theta_{0})\neq0$ rejects the null if $\mathcal{W}_{MR}>z^{*}_{\mathcal{W},\alpha}$. Similarly, each of the following bootstrap CI's are of asymptotic confidence level $1-\alpha$. The lower endpoint one-sided bootstrap CI is $[\hat{\theta}_{r}- z_{T,\alpha}^{*}\sqrt{\hat{\Sigma}_{MR,r}/n},\infty)$ which corresponds to the one-sided bootstrap $t$ test above. The symmetric and the equal-tailed bootstrap percentile-$t$ intervals are $[\hat{\theta}_{r}\pm z^{*}_{|T|,\alpha}\sqrt{\hat{\Sigma}_{MR,r}/n}]$ and $[\hat{\theta}_{r}- z^{*}_{T,\alpha/2}\sqrt{\hat{\Sigma}_{MR,r}/n},\hat{\theta}_{r}- z^{*}_{T,1-\alpha/2}\sqrt{\hat{\Sigma}_{MR,r}/n}]$\footnote{The formula may look confusing to readers. It is correct that the upper end of the CI is $\hat{\theta}_{r}- z^{*}_{T,1-\alpha/2}\sqrt{\hat{\Sigma}_{MR,r}/n}$, not $\hat{\theta}_{r}+ z^{*}_{T,1-\alpha/2}\sqrt{\hat{\Sigma}_{MR,r}/n}$ because $z^{*}_{T,1-\alpha/2}<z^{*}_{T,\alpha/2}$.}, respectively. The Wald-based bootstrap confidence region for $\eta(\theta_{0})$ is $\{\eta\in\mathbf{R}^{L_{\eta}}:n\cdot (\eta(\hat{\theta})-\eta)'((\partial\eta(\hat{\theta})/\partial\theta')\hat{\Sigma}_{MR}(\partial\eta(\hat{\theta})/\partial\theta')')^{-1}(\eta(\hat{\theta})-\eta)\leq z^{*}_{\mathcal{W},\alpha}\}$.

In sum, the misspecification-robust bootstrap procedure is as follows:
\begin{enumerate}
  \itemsep0em 
  \item Draw $n$ random observations $\chi_{n}^{*}$ with replacement from the original sample, $\chi_{n}$.
  \item Calculate $\hat{\theta}^{*}$ and $\hat{\Sigma}^{*}_{MR}$ using the same formula with their sample counterparts.
  \item Construct and save $T_{MR}^{*}$ or $\mathcal{W}^{*}_{MR}$.
  \item Repeat steps 1-3 $B$ times and get the distribution of $T_{MR}^{*}$ or $\mathcal{W}^{*}_{MR}$.
  \item Find $z^{*}_{|T|,\alpha}$, $z^{*}_{T,\alpha}$, or $z^{*}_{\mathcal{W},\alpha}$ from the distribution of $|T_{MR}^{*}|$,  $T_{MR}^{*}$, or $\mathcal{W}^{*}_{MR}$.
\end{enumerate}

\section{Main Result}
\label{S_Main}

Let $f(X_{i},\beta)$ be a vector containing the unique components of $\psi(X_{i},\beta)$ and its derivatives with respect to the components of $\beta$ through order $d$, and $\psi(X_{i},\beta)\psi(X_{i},\beta)'$ and its derivatives with respect to the components of $\beta$ through order $d-1$. 

\begin{assumption}
$X_{i},i=1,2,...n$ are iid.
\label{A1}
\end{assumption}

\begin{assumption}\
	\vspace{-0.5em}
\begin{description}
  \itemsep0em 
  \item[(a)] $\Theta$ is a compact parameter space of $\theta$ such that $\theta_{0}$ is an interior point of $\Theta$; $\Lambda(\theta)$ is a compact parameter space of $\lambda(\theta)$ such that it contains a zero vector and $\lambda_{0}(\theta)$ is an interior point of $\Lambda(\theta)$.
  \item[(b)] $(\hat{\theta},\hat{\lambda})$ solves \eqref{GEL} for EL or ET, or \eqref{ETEL} for ETEL; $(\theta_{0},\lambda_{0})$ is the pseudo-true value that uniquely solves the population version of \eqref{GEL} for EL or ET, or \eqref{ETEL} for ETEL.
  \item[(c)] For some function $C_{g}(x)$, $\|g(x,\theta_{1})-g(x,\theta_{2})\|<C_{g}(x)\|\theta_{1}-\theta_{2}\|$ for all $x$ in the support of $X_{1}$ and all $\theta_{1},\theta_{2}\in\Theta$; $EC_{g}^{q_{g}}(X_{1})<\infty$ and $E\|g(X_{1},\theta)\|^{q_{g}}<\infty$ for all $\theta\in\Theta$ for all $0<q_{g}<\infty$.
  \item[(d)] For some function $C_{\rho}(x)$, $|\rho(\lambda_{1}'g(x,\theta_{1}))-\rho(\lambda_{2}'g(x,\theta_{2}))|<C_{\rho}(x)\|(\theta_{1}',\lambda_{1}')-(\theta_{2}',\lambda_{2}')\|$ for all $x$ in the support of $X_{1}$, all $\theta_{l}\in\Theta$, and all $\lambda_{l}\in\Lambda(\theta_{l})$ for $l=1,2$; $EC_{\rho}^{q_{1}}(X_{1})<\infty$ for some $q_{1}>4$. In addition, UBC \eqref{UBC} holds for EL. 
\end{description}
\label{A2}
\end{assumption}

\begin{assumption}\
	\vspace{-0.5em}
\begin{description}
  \itemsep0em 
  \item[(a)] $\Gamma$ is nonsingular and $\Psi$ is positive definite. 
  \item[(b)] $g(x,\theta)$ is $d+1$ times differentiable with respect to $\theta$ on $N(\theta_{0})$, some neighborhood of $\theta_{0}$, for all $x$ in the support of $X_{1}$, where $d\geq 4$.
  \item[(c)] There is a function $C_{G}(x)$ such that $\|G^{(j)}(x,\theta)-G^{(j)}(x,\theta_{0})\|\leq C_{G}(x)\|\theta-\theta_{0}\|$ for all $x$ in the support of $X_{1}$ and all $\theta\in N(\theta_{0})$ for $j=0,1,...,d+1$; $EC_{G}^{q_{G}}(X_{1})<\infty$ and $E\|G^{(j)}(X_{1},\theta_{0})\|^{q_{G}}<\infty$ for $j=0,1,...,d+1$ for all $0<q_{G}<\infty$.
  \item[(d)] There is a function $C_{\partial\rho}(x)$ such that \[|\rho_{j}(\lambda'g(x,\theta))-\rho_{j}(\lambda_{0}'g(x,\theta_{0}))|\leq C_{\partial\rho}(x)\|(\theta',\lambda')-(\theta_{0}',\lambda_{0}')\|\] for all $x$ in the support of $X_{1}$, all $\lambda\in\Lambda(\theta)$, all $\theta\in N(\theta_{0})$ for $j=1,...,d+1$; $EC_{\partial\rho}^{q_{2}}(X_{1})<\infty$ for some $q_{2}>16$.
  \item[(e)] $f(X_{1},\beta_{0})$ is once differentiable with respect to $X_{1}$ with uniformly continuous first derivative.
  \item[(f)] For the Wald statistic, the $\mathbf{R}^{L_{\eta}}$-valued function $\eta(\cdot)$ is $d$ times continuously differentiable at $\theta_{0}$ and $(\partial/\partial\theta')\eta(\theta_{0})$ is full rank $L_{\eta}\leq L_{\theta}$.
\end{description}
\label{A3}
\end{assumption}

\begin{assumption}
For $t\in\mathbf{R}^{dim(f)}$, $\limsup_{\|t\|\rightarrow\infty}\left|Ee^{it'f(X_{1},\beta_{0})}\right|<1,$ where $i=\sqrt{-1}$.
\label{A4}
\end{assumption}

Assumption \ref{A1} is that the sample is iid, which is also assumed in Schennach (2007) and Newey and Smith (2004). Assumption \ref{A2}(a)-(c) are similar to Assumption 2(a)-(b) of Andrews (2002). Assumption \ref{A2}(d) is similar to but slightly stronger than Assumption 3(4) of Schennach (2007) for ET or ETEL, and it includes Assumption 3(1) of Chen, Hong, and Shum (2007) for EL to avoid a negative implied probability under misspecification. Assumption \ref{A2}(c)-(d) are required to have the uniform convergence of the objective function. Assumption \ref{A3}(a) is a standard regularity condition for a well-defined asymptotic covariance matrix. Assumption \ref{A3} except for (d) is similar to Assumption 3 of Andrews (2002). The assumptions on $q_{g}$ and $q_{G}$ are slightly stronger than necessary, but yield a simpler result. This is also assumed in Andrews (2002) for the same reason. Assumption \ref{A3}(d) is similar to but stronger than Assumption 3(6) of Schennach (2007). It ensures that the components of higher-order Taylor expansion of the FOC have well-defined probability limits.\footnote{The values of $q_{g}$, $q_{G}$, $q_{1}$, and $q_{2}$ are determined to ensure the existence of higher-order moments which is required for asymptotic refinements through Edgeworth expansions. Consistency of the bootstrap, however, can be shown under weaker assumptions.} Assumption \ref{A4} is the standard Cram\'{e}r condition for Edgeworth expansion, and it is satisfied if the distribution of $f(X_{1},\beta_{0})$ has a probability density with respect to Lebesgue measure (Horowitz, 2001).

Theorem \ref{T1} formally establishes asymptotic refinements of the bootstrap $t$ and Wald tests based on EL, ET, and ETEL estimators. This result is new, because asymptotic refinements of the bootstrap for this class of estimators have not been established in the literature even under correct model specifications.

\begin{theorem}
(a) Suppose Assumptions \ref{A1}-\ref{A4} hold with $q_{1}>4$ and $q_{2}>\frac{16}{1-2\xi}$ for some $\xi\in[0,1/2)$. Under $H_{0}:\theta_{r}=\theta_{0,r}$,
\begin{eqnarray}
\nonumber && P(T_{MR}>z^{*}_{T,\alpha}) = \alpha + o(n^{-(1/2+\xi)}) \text{ and}\\
\nonumber && P(T_{MR}<z^{*}_{T,\alpha/2} \text{ or }T_{MR}>z^{*}_{T,1-\alpha/2})=\alpha+o(n^{-(1/2+\xi)}).
\end{eqnarray}
(b) Suppose Assumptions \ref{A1}-\ref{A4} hold with $q_{1}>6$, $q_{2}>\frac{30}{1-2\xi}$ for some $\xi\in[0,1/2)$, and $d\geq5$. Under $H_{0}:\theta_{r}=\theta_{0,r}$,
\[P(|T_{MR}|>z^{*}_{|T|,\alpha})=\alpha+o(n^{-(1+\xi)}).\]
Under $H_{0}:\eta(\theta_{0})=0$,
\[P(\mathcal{W}_{MR}>z^{*}_{\mathcal{W},\alpha})=\alpha+o(n^{-(1+\xi)}).\]
(c) Suppose Assumptions \ref{A1}-\ref{A4} hold with $q_{1}>8$, a sufficiently large $q_{2}$, and $d\geq6$. Under $H_{0}:\theta_{r}=\theta_{0,r}$,
\[P(|T_{MR}|>z^{*}_{|T|,\alpha})=\alpha+O(n^{-2}).\]
\label{T1}
\end{theorem}
\vspace{-2em}
\begin{remark}[Remark 1]
By the duality of $t$ tests and CI's, asymptotic refinements of the same rate for the bootstrap CI's follow from Theorem \ref{T1}. The equal-tailed percentile-$t$ CI corresponds to Theorem \ref{T1}(a). The symmetric percentile-$t$ CI corresponds to Theorem \ref{T1}(b)-(c). The Wald confidence region corresponds to Theorem \ref{T1}(b). Recall that the asymptotic $t$, Wald tests, and CI's based on $T_{MR}$ and $\mathcal{W}_{MR}$ are correct up to $O(n^{-1/2})$, $O(n^{-1})$, and $O(n^{-1})$ for (a), (b), and (c), respectively. The two-sided bootstrap $t$ and Wald tests, and the symmetric percentile-$t$ CI achieve a higher rate of refinements because the $O(n^{-1/2})$ terms of the Edgeworth expansions of the corresponding statistics are zero by a symmetry property.
\end{remark}
\begin{remark}[Remark 2]
The result in Theorem \ref{T1}(c) is sharp and based on the argument of Hall (1988). By using the Edgeworth and Cornish-Fisher expansions, Hall showed that the $O(n^{-3/2})$ term of the coverage probability of the symmetric percentile-$t$ CI is also zero. Since his derivation is based on the one-dimensional $t$ statistic, I do not formally state a similar result for the Wald test. However, it is likely that the same sharp rate of refinements would hold (e.g. Hall, 1992, Section 4.2; Horowitz, 2001, Section 3.3). 
\end{remark}

The proof of Theorem \ref{T1} follows the steps of Andrews (2002) that establish asymptotic refinements of the bootstrap for GMM estimators under correct specification. I briefly outline the proof. The conclusion of Theorem \ref{T1}(a) follows from
\begin{equation}
\label{outline1}
P\left(\sup_{z\in\mathbf{R}}\left|P^{*}(T_{MR}^{*}\leq z)-P(T_{MR}\leq z)\right|>n^{-(1/2+\xi)}\varepsilon\right)=o(n^{-1})
\end{equation}
for any $\varepsilon>0$, which leads to 
\begin{equation}
\label{outline2}
1-\alpha-n^{-(1/2+\xi)}\varepsilon+o(n^{-1})\leq P(T_{MR}\leq z^{*}_{T,\alpha}) \leq 1-\alpha+n^{-(1/2+\xi)}\varepsilon+o(n^{-1}).
\end{equation}
Since the $o(n^{-1})$ terms in \eqref{outline2} are directly related to the $o(n^{-1})$ term on the right-hand side (RHS) of \eqref{outline1}, it is critical to show in \eqref{outline1} that the random cdf $P^{*}(T^{*}_{MR}\leq z)$ differs from the nonrandom cdf $P(T_{MR}\leq z)$ by a small amount, $n^{-(1/2+\xi)}$, on a set with probability $o(n^{-1})$, rather than $o(1)$. Similar arguments apply to the conclusions of Theorem \ref{T1}(b)-(c). \eqref{outline1} is shown by using Hall (1992)'s argument on Edgeworth expansion of a smooth function of sample averages. That is, I show that $T_{MR}$ and $T_{MR}^{*}$ are well approximated by a smooth function of the sample and the bootstrap sample moments (Lemma \ref{L7}), and the smooth function allows Edgeworth expansions up to a certain order (Lemma \ref{L9}). The argument of the smooth function consists of the elements of $n^{-1}\sum_{i=1}^{n}f(X_{i},\beta_{0})$ and $n^{-1}\sum_{i=1}^{n}f(X_{i}^{*},\hat{\beta})$, whose consistency is shown in Lemma \ref{L6}. The components of the Edgeworth expansions are well defined and consistent (Lemma \ref{L8}). Lemmas \ref{L2}-\ref{L5} establish consistency of the sample and the bootstrap GEL estimators.

Since I derive the asymptotic distribution of GEL under misspecification by using the fact that GEL FOC forms a just-identified GMM, one might wonder why the proof is different from that of GMM. The proof of this paper can be divided into two parts: (i) consistency, and (ii) higher-order analysis, and each part is a nontrivial extension. First, consistency of GEL estimators should be shown for the solution to the original GEL minimax criterion as the FOC can have multiple roots even when the original minimax criterion has a unique solution (e.g. Newey and McFadden, 1994, p. 2117). Additional complications arise due to the fact that we require a stronger result than usual consistency to control the error in the bootstrap approximation as in \eqref{outline1}. Thus, the consistency proof of this paper cannot be simplified to that of a just-identified GMM. Second, an economical higher-order analysis is required because the existence and finiteness of GEL higher-order moments may restrict model misspecification. Although it is commonly assumed that all of higher-order moments of GMM moment function and its higher-order derivatives are finite (Andrews, 2002), this does not affect robustness to misspecification of the bootstrap (Lee, 2014). However, this conclusion cannot be directly applied to GEL, because GEL FOC, which forms a just-identified GMM moment function, contains the Lagrange multiplier $\hat{\lambda}$ whose probability limit is zero under correct specification but is non-zero under misspecification. For example, suppose that Assumptions \ref{A2}(d) and \ref{A3}(d) hold for all $0<q_{l}<\infty$ for $l=1,2$, that would be assumed if we naively mapped the assumptions of GMM onto GEL. Since a zero vector is in $\Lambda(\theta)$, this implies $Ee^{q_{l}\lambda_{0}'g(X_{i},\theta_{0})}<\infty$ for ET, which is a strong assumption on DGP and the model. Since $\lambda_{0}\neq 0$ under misspecification, $Ee^{q_{l}\lambda_{0}'g(X_{i},\theta_{0})}$ may not be finite if $q_{l}$ is too large, and the bootstrap would not achieve desired asymptotic refinements. Lee (2014b) provides an example that the model cannot be too misspecified to satisfy the assumptions and the set of possible misspecification shrinks to zero as $q_{l}$ gets larger. This implies that by assuming $0<q_{l}<\infty$, one may completely rule out model misspecification. Thus, it is important to find stringent conditions for $q_{1}$ and $q_{2}$, and this requires an analysis of GEL higher-order moments and their higher-order derivatives. 

\section{Monte Carlo Results}
\label{S_MC}

This section compares the finite sample CI coverage probabilities under correct specification and misspecification. To reduce computational burden of calculating GEL estimators $B$ times for each Monte Carlo repetition, the warp-speed Monte Carlo method of Giacomini, Politis, and White (2013) is used. The method also appears in White (2000) and Davidson and MacKinnon (2002, 2007), but the validity of the method is formally established by Giacomini, Politis, and White. The key difference between the warp-speed method and a usual Monte Carlo is that the bootstrap sample is drawn only once for each Monte Carlo repetition rather than $B$ times, and thus computation time is significantly reduced. The number of Monte Carlo repetition is 5,000 throughout this section. I consider the AR(1) dynamic panel model of Blundell and Bond (1998). For $i=1,...,n$ and $t=1,...T$,
\begin{eqnarray}
y_{it} &=& \rho_{0} y_{i,t-1} + \eta_{i} + \nu_{it},
\label{AR1}
\end{eqnarray}
where $\eta_{i}$ is an unobserved individual-specific effect and $\nu_{it}$ is an error term. To estimate $\rho_{0}$, we use two sets of moment conditions: 
\begin{eqnarray}
Ey_{i(t-s)}(\Delta y_{it}-\rho_{0}\Delta y_{i(t-1)}) = 0,&& \hspace{0.5em}t=3,...T, \text{ and }s\geq 2,
\label{DIF}\\
E\Delta y_{i(t-1)}(y_{it}-\rho_{0}y_{i(t-1)}) = 0,&& \hspace{0.5em}t=3,...T.
\label{SYS}
\end{eqnarray}
The first set \eqref{DIF} is derived from taking differences of \eqref{AR1}, and uses the lagged values of $y_{it}$ as instruments. The second set \eqref{SYS} is derived from the initial conditions on DGP and mitigates the weak instruments problem from using only the lagged values. Blundell and Bond (1998) suggest to use the system-GMM estimator based on the two sets of moment conditions. The number of moment conditions is $(T+1)(T-2)/2$.  

Four DGP's are considered: two correctly specified and two misspecified models. For each of the DGP's, $T=4, 6$ and $n=100, 200$ are considered. To minimize the effect of the initial condition, I generate 100+T time periods and use the last T periods for estimation. In Tables \ref{table_C1}-\ref{table_M2}, ``Boot'' and ``Asymp'' mean the bootstrap CI and the asymptotic CI, respectively. The third column shows on which estimator the CI is based. GMM denotes the two-step GMM based on the system moment conditions. The fourth column shows which standard error is used: ``C'' denotes the conventional standard error and ``MR'' denotes the misspecification-robust one. The fifth column shows how the bootstrap is implemented for the bootstrap CI's: ``L'' denotes the misspecification-robust bootstrap proposed in this paper and in Lee (2014), ``HH'' denotes the recentering method of Hall and Horowitz (1996), and ``BN'' denotes the efficient bootstrapping of Brown and Newey (2002). The columns under ``CI'' show the coverage probabilities. The column under ``J test'' shows the rejection probabilities of the overidentification tests: the Hall-Horowitz bootstrap J test, the asymptotic J test, the likelihood-ratio tests based on EL, ET, and ETEL, are presented.

In sum, eight bootstrap CI's and eight asymptotic CI's are compared. Boot-GMM-C-HH serves as a benchmark, as its properties have been relatively well investigated. Boot-GMM-MR-L is suggested by Lee (2014). The theoretical advantage of Boot-EL-MR-L is established in this paper. Boot-EL-MR-BN uses the EL probabilities in resampling. This paper does not establish asymptotic refinements for this CI and the efficient resampling method (BN) is not robust to misspecification as is discussed in Section 2. However, its performance is worth attention and the efficient resampling may be modified using shrinkage to make the CI robust to misspecification. CI's based on ET and ETEL are defined similarly. Note that CI's using the conventional standard error (C), either bootstrap or asymptotic, are not robust to misspecification. 

The DGP for a correctly specified model is identical to that of Bond and Windmeijer (2005). For $i=1,...,n$ and $t=1,...T$,
\begin{eqnarray}
\nonumber \text{DGP C-1: }&& y_{it} = \rho_{0} y_{i,t-1} + \eta_{i} + \nu_{it},\\
\nonumber && \eta_{i}\sim N(0,1);\nu_{it}\sim \frac{\chi_{1}^{2}-1}{\sqrt{2}},\\
\nonumber && y_{i1} = \frac{\eta_{i}}{1-\rho_{0}}+u_{i1};u_{i1}\sim N\left(0,\frac{1}{1-\rho_{0}^{2}}\right).
\end{eqnarray} 
Since the bootstrap does not solve weak instruments (Hall and Horowitz, 1996), I let $\rho_{0}=0.4$ so that the performance of the bootstrap is not affected by the problem. The simulation result is given in Table \ref{table_C1}. First of all, the bootstrap CI's show significant improvement over the asymptotic CI's across all the cases considered. Second, similar to the result of Bond and Windmeijer (2005), the bootstrap CI's coverage probabilities tend to be too high for $T=6$. This over-coverage problem becomes less severe as the sample size increases, especially for those based on EL, ET, and ETEL. Interestingly, efficient resampling (BN) seems to mitigate this problem. Third, the asymptotic CI's using the robust standard error (MR) work better than the ones using the usual standard error (C). This result is surprising given that the model is correctly specified. One reason is that both standard errors underestimate the standard deviation of the estimator while the robust standard error is relatively large in this case. For example, when $T=6$ and $n=100$, the difference in the coverage probabilities between Asymp-ET-C and Asymp-ET-MR is quite large. The unreported standard deviation of the ET estimator is .085, while the mean of the robust and the conventional standard errors are .059 and .047, respectively. Finally, the overidentification tests except for the asymptotic J test show severe size distortion, especially when $T=6$. 

Next a heteroskedastic error term across individuals is considered. The DGP is
\begin{eqnarray}
\nonumber \text{DGP C-2: }&& y_{it} = \rho_{0} y_{i,t-1} + \eta_{i} + \nu_{it},\\
\nonumber && \eta_{i}\sim N(0,1);\nu_{it}\sim N(0,\sigma_{i}^{2});\sigma_{i}^{2}\sim U[0.2.1.8],\\
\nonumber && y_{i1} = \frac{\eta_{i}}{1-\rho_{0}}+u_{i1};u_{i1}\sim N\left(0,\frac{\sigma_{i}^{2}}{1-\rho_{0}^{2}}\right).
\end{eqnarray} 
The result is given in Table \ref{table_C2}. The findings are similar to that of Table \ref{table_C1}, except that the over-coverage problem of the bootstrap CI's based on GEL estimators improves more quickly as the sample size grows.

To allow misspecification, suppose that the DGP follows an AR(2) process while the model is based on the AR(1) specification, \eqref{AR1}. For $i=1,...,n$ and $t=1,...T$,
\begin{eqnarray}
\nonumber \text{DGP M-1: }&& y_{it} = \rho_{1} y_{i,t-1} + \rho_{2} y_{i,t-2} + \eta_{i} + \nu_{it},\\
\nonumber && \eta_{i}\sim \text{tr}N(0,1);\nu_{it}\sim \frac{\text{tr}\chi_{1}^{2}-1}{\sqrt{2}},\\
\nonumber && y_{i1} = \frac{\eta_{i}}{1-\rho_{1}-\rho_{2}}+u_{i1};u_{i1}\sim  \sqrt{\frac{1-\rho_{2}}{(1+\rho_{2})[(1-\rho_{2})^{2}-\rho_{1}^{2}]}}\cdot \text{tr}N\left(0,1\right),
\end{eqnarray}
where $\text{tr}N(0,1)$ and $\text{tr}\chi_{1}^{2}$ are truncated standard normal between -4 and 4, and truncated chi-square distribution with 1 degrees of freedom between 0 and 16, respectively. The truncated distributions are used to satisfy the UBC \eqref{UBC}. DGP M-2 is identical to DGP M-1 except that $\nu_{it}$ is truncated log normal distribution between $-\sqrt{e}$ and $e^{3.5}$ with mean zero. 

If the model is misspecified, then there is no true parameter that satisfies the moment conditions simultaneously. It is important to understand what is identified and estimated under misspecification. The moment conditions \eqref{DIF} and \eqref{SYS} impose 
\begin{equation}
\frac{Ey_{i1}\Delta y_{it}}{Ey_{i1}\Delta y_{i(t-1)}} = \cdots = \frac{Ey_{i(t-3)}\Delta y_{it}}{Ey_{i(t-3)}\Delta y_{i(t-1)}} = \frac{Ey_{i(t-2)}\Delta y_{it}}{Ey_{i(t-2)}\Delta y_{i(t-1)}}=\frac{E\Delta y_{i(t-1)} y_{it}}{E\Delta y_{i(t-1)} y_{i(t-1)}},
\label{MCS}
\end{equation}
for $t=3,...,T$. Under correct specification, the restriction \eqref{MCS} holds and a unique parameter is identified. However, each of the ratios identifies different parameters under misspecification, and the probability limits of GMM and GEL estimators are weighted averages of the parameters. For example, when $T=4$, we have five moment conditions. Four of them identify $\rho_{T4}^{a}\equiv\rho_{1}-\rho_{2}$ and the other identify $\rho_{T4}^{b}\equiv\rho_{1} + \frac{\rho_{2}}{\rho_{1}-\rho_{2}}$. Thus, the pseudo-true value $\rho_{0}$ is defined as $\rho_{0}=w\rho_{T4}^{a}+(1-w)\rho_{T4}^{b}$ where $w$ is between 0 and 1. Similarly, the pseudo-true value when $T=6$ is a weighted average of four different parameters. Since GMM and GEL use different weights, the pseudo-true values would be different. If $\rho_{2}=0$, then the pseudo-true values coincide with $\rho_{1}$, the AR(1) coefficient. Thus, $\rho_{0}$ captures the deviation from the AR(1) model. If $|\rho_{2}|$ is relatively small, then $\rho_{0}$ would not be much different from $\rho_{1}$, while there is an advantage of using a parsimonious model. If one accepts the possibility of misspecification and decides to proceed with the pseudo-true value, then GEL pseudo-true values have better interpretation than GMM ones because GEL weights are implicitly calculated according to a well-defined distance measure while GMM weights depend on the choice of a weight matrix by a researcher.

Tables \ref{table_M1}-\ref{table_M2} show the CI coverage probabilities under DGP M-1 and M-2, respectively. I set $\rho_{1}=0.6$ and $\rho_{2}=0.2$. The pseudo-true values are calculated using the sample size of $n=30,000$ for $T=4$ and $n=20,000$ for $T=6$.\footnote{The two-step GMM and GEL pseudo-values are not that different. They are around 0.4 when $T=4$ and around 0.5 when $T=6$.} It is clearly seen that the bootstrap CI's outperform the asymptotic CI's. In particular, the performances of Boot-EL-MR-L, Boot-ET-MR-L, and Boot-ETEL-MR-L CI's are excellent for $T=4$. When $T=6$, these CI's exhibit slight over-coverage but less severe than Boot-GMM-MR-L.\footnote{Observing that the over-coverage problem of the bootstrap CI's becomes severe as $T$ gets larger, I conjecture that this problem is related to the estimation of the misspecification-robust variance matrix because the dimension of the matrix increases along with $T$.} The bootstrap CI's using the efficient resampling (BN) show some improvement on the over-coverage problem, but they are not robust to misspecification. Indeed, their coverage probabilities deviate from the nominal ones as the sample size grows in DGP M-2. One may wonder why the HH bootstrap CI works quite well under misspecification even though the CI is not robust to misspecification. This is spurious and cannot be generalized. In this case, the conventional standard error is considerably smaller than the robust standard error, while the HH bootstrap critical value is much larger than the asymptotic one, which offsets the smaller standard error. Lee (2014) reports that the performance of the HH bootstrap CI under misspecification is much worse than that of the MR bootstrap CI. In addition, the HH bootstrap J test shows very low power relative to the asymptotic tests. Among the asymptotic CI's, those based on GEL estimators and the robust standard errors (MR) show better performances. 

Finally, Table \ref{table_W1} compares the width of the bootstrap CI's under different DGP's. Since this paper establishes asymptotic refinements in the size and coverage errors, the width of CI's is not directly related to the main result. Nevertheless, the table clearly demonstrates a reason to consider GEL as an alternative to GMM, especially when misspecification is suspected. Under correct specification (C-1 and C-2), all the bootstrap CI's have similar width. This conclusion changes dramatically under misspecification (M-1 and M-2). Among robust CI's, (Boot-)GMM-MR-L is much wider than those based on GEL. For example, when $T=4$ and $n=200$ in DGP M-1, the width of the (Boot-)GMM-MR-L 95\% CI is 2.418, while that of (Boot-)ETEL-MR-L 95\% CI is only .880. The main reason is that the GEL standard errors are smaller than the GMM ones under misspecification, at least for the considered DGP's. The bootstrap CI's using the efficient resampling (BN) are generally narrower than those using the iid resampling. This suggests that the GEL probabilities with appropriate shrinkage may be used to shorten CI's under misspecification.

The findings of Monte Carlo experiments can be summarized as follows. First, the misspecification-robust bootstrap CI's based on GEL estimators are generally more accurate than other bootstrap and asymptotic CI's regardless of misspecification. Not surprisingly, the coverage of non-robust CI's are very poor under misspecification. Second, the GEL-based bootstrap CI's improve on the severe over-coverage of the GMM-based bootstrap CI's, which is also a concern of Bond and Windmeijer (2005). Lastly, it is recommended to use the misspecification-robust variance estimator in constructing $t$ statistics and CI's regardless of whether the model is correctly specified or not, because the coverage of the misspecification-robust CI's tends to be more accurate even under correct specification.

\section{Application: Returns to Schooling}
\label{S_HI}

Hellerstein and Imbens (1999) estimate the Mincer equation by weighted least squares, where the weights are calculated using EL. The equation of interest is 
\begin{eqnarray}
\nonumber \log(\text{wage}_{i}) &=& \beta_{0} + \beta_{1}\cdot \text{education}_{i} + \beta_{2}\cdot \text{experience}_{i} + \beta_{3} \cdot \text{experience}^{2}_{i}\\
&& + \beta_{4}\cdot \text{IQ}_{i} + \beta_{5} \cdot \text{KWW}_{i} + \varepsilon_{i},
\label{Mincer}
\end{eqnarray}
where KWW denotes Knowledge of the World of Work, an ability test score. Since the National Longitudinal Survey Young Men's Cohort (NLS) dataset reports both ability test scores and schooling, the equation \eqref{Mincer} can be estimated by OLS. However, the NLS sample size is relatively small, and it may not correctly represent the whole population. In contrast, the Census data is a very large dataset which is considered as the whole population, but we cannot directly estimate the equation \eqref{Mincer} using the Census because it does not contain ability measures. Hellerstein and Imbens calculate weights by matching the Census and the NLS moments and use the weights to estimate the equation \eqref{Mincer} by the least squares. This method can be used to reduce the standard errors or change the estimand toward more representative of the Census. 

Let $y_{i}\equiv \log(\text{wage}_{i})$ and $\mathbf{x}_{i}$ be the regressors on the right-hand-side of \eqref{Mincer}. The Hellerstein-Imbens weighted least squares can be viewed as a special case of the EL estimator using the following moment condition:
\begin{equation}
E_{s}g_{i}(\beta_{0})=0,
\label{mc}
\end{equation}
where $E_{s}[\cdot]$ is the expectation over a probability density function $f_{s}(y_{i},\mathbf{x}_{i})$, which is labeled the \textit{sampled population}. The moment function $g_{i}(\beta)$ is 
\begin{equation}
g_{i}(\beta) = \left(\begin{array}{c} \mathbf{x}_{i}(y_{i}-\mathbf{x}_{i}'\beta) \\ m(y_{i},\mathbf{x}_{i}) - E_{t}m(y_{i},\mathbf{x}_{i})\end{array}\right),
\label{mf}
\end{equation}
where $\beta$ is a parameter vector, $m(y_{i},\mathbf{x}_{i})$ is a $13 \times 1$ vector, and $E_{t}[\cdot]$ is the expectation over a probability density function $f_{t}(y_{i},\mathbf{x}_{i})$, labeled the \textit{target population}. The first set of the moment condition is the FOC of OLS and the second set matches the sample (NLS) moments with the known population (Census) moments. In particular, the thirteen moments consisting of first, second, and cross moments of log(wage), education, experience, and experience squared are matched. If the sampled population is identical to the target population, i.e., the NLS sample is randomly drawn from the Census distribution, the moment condition model is correctly specified and \eqref{mc} holds. Otherwise, the model is misspecified and there is no such $\beta$ that satisfies \eqref{mc}. In this case, the probability limit of the EL estimator solves the FOC of OLS with respect to an artificial population that minimizes a distance between the sampled and the target populations. This pseudo-true value is an interesting estimand because we are ultimately interested in the parameters of the target population, rather than the sampled population.

Table \ref{tb_est} shows the estimation result of OLS, two-step GMM, EL, ET, and ETEL estimators. Without the Census moments, the equation \eqref{Mincer} is estimated by OLS and the estimate of the returns to schooling is 0.054 with the standard error of 0.010. By using the Census moments, the coefficients estimates and the standard errors change. The two-step GMM estimator is calculated using the OLS estimator as a preliminary estimator, and it serves as a benchmark. EL, ET, and ETEL produce higher point estimates and smaller standard errors than those of OLS. Since the J-test rejects the null hypothesis of correct specification for all of the estimators using the Census moments, it is likely that the target population differs from the sampled population. If this is the case, then the conventional standard errors are no longer valid, and the misspecification-robust standard errors should be used. The misspecification-robust standard errors, s.e.$_{MR}$, of EL, ET, and ETEL are slightly larger than the usual standard errors assuming correct specification, s.e.$_{C}$, but still smaller than the standard errors of OLS. In contrast, s.e.$_{MR}$ of GMM is much larger than s.e.$_{C}$, which is consistent with the simulation result given in Section \ref{S_MC}.

\begin{figure}[p!]
	\centering
	\includegraphics[width=140mm]{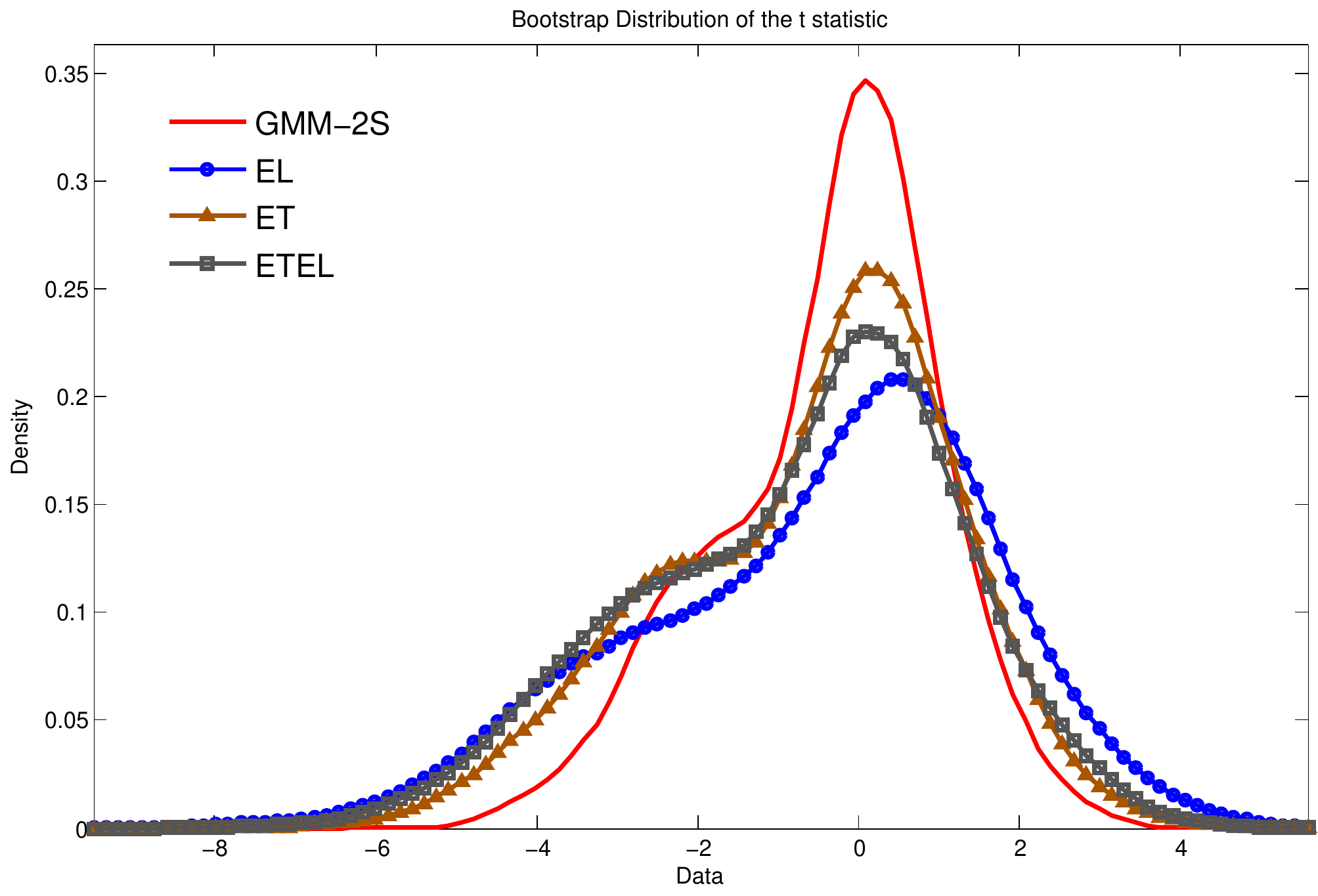}
	\caption{Bootstrap distribution of the $t$ statistics based on 2-step GMM estimator (solid), EL estimator (with circle), ET estimator (with triangle), and ETEL estimator (with rectangle).}
	\label{fig_bootdist}
\end{figure}

Table \ref{tb_CI} shows the lower and upper bounds of CI's based on various estimators and their respective width. The width of the GMM based CI's are wider than those based on GEL estimators. Among the GEL estimators, the ET estimator has the widest CI, while the EL estimator has the narrowest. The asymptotic CI's are narrower than the bootstrap CI's, but this is likely to cause under-coverage given the simulation result in Section \ref{S_MC}. The upper bounds of the bootstrap CI's range from 9.6\% to 11.5\%, which are higher than those of the asymptotic CI's. I also present a nonparametric kernel estimate of the bootstrap distribution of the $t$ statistics based on GMM, EL, ET, and ETEL estimators in Figure \ref{fig_bootdist}. The distributions are skewed to the left, which implies the presence of a downward bias. Overall, the estimation of \eqref{Mincer} using GEL estimators and the resulting bootstrap CI's suggest that the returns to schooling is likely to be higher than originally estimated by Hellerstein and Imbens.

\section{Conclusion}
\label{S_Con}

GEL estimators are favorable alternatives to GMM. Although asymptotic refinements of the bootstrap for GMM have been established, the same for GEL have not been done yet. In addition, the current literature on bootstrapping does not consider model misspecification that adversely affects the refinement and validity of the bootstrap. This paper formally established asymptotic refinements of the bootstrap for $t$ and Wald tests, and CI's and confidence regions based on GEL estimators. Moreover, the proposed bootstrap is robust to misspecification, which means the refinements are not affected by model misspecification. Simulation results did support this finding. As an application, the returns to schooling was estimated by extending the method of Hellerstein and Imbens (1999). The exercise found that the estimates of Hellerstein and Imbens were robust across different GEL estimators, and the returns to schooling could be even higher.

\section*{Acknowledgment}
I am very grateful to Bruce Hansen and Jack Porter for their encouragement and helpful comments. I also thank the co-editor Han Hong, an associate editor, three anonymous referees, Guido Imbens, Xiaohong Chen, and Yoon-Jae Whang, as well as seminar participants at UW-Madison, Monash, ANU, Adelaide, UNSW, and U of Sydney for their suggestions and comments. This paper was also presented at the 2013 NASM and SETA 2013.

\appendix
\allowdisplaybreaks
\section{Appendix: Lemmas and Proofs}

\subsection{Proof of Proposition 1}

\textit{proof.} The proof is similar to that of Theorem 10 of Schennach (2007), and thus omitted.

\subsection{Lemmas}

The lemmas and proofs extensively rely on Hall and Horowitz (1996), Andrews (2002), and Schennach (2007). For brevity, Hall and Horowitz (1996) is abbreviated to HH, Andrews (2002) to A2002, and Schennach (2007) to S2007. In particular, I use Lemmas 1, 2, 6, and 7 of A2002 with minor modifications for a nonparametric iid bootstrap. They are denoted by AL1, AL2, AL6, and AL7, respectively. Such modifications are justified by Lemma 1 of Lee (2014), which holds under our Assumptions 1-3. In addition, Lemma 5 of A2002 is denoted by AL5 without modification.

\vspace{1em}
Lemma \ref{L2} shows the uniform convergence of the so-called inner loop and the objective function in $\theta$. Since ET and ETEL solve the same inner loop optimization problem, we let $\rho(\nu)=1-e^{\nu}$ for ETEL in the next lemma. Define $\hat{\lambda}(\theta)=\argmax_{\lambda\in\mathbf{R}^{L_{g}}} n^{-1}\sum_{i=1}^{n}\rho(\lambda'g_{i}(\theta))$ and $\lambda_{0}(\theta)=\argmax_{\lambda\in\mathbf{R}^{L_{g}}} E\rho(\lambda'g_{i}(\theta))$. Such solutions exist and are continuously differentiable around a neighborhood of $\hat{\theta}$ and $\theta_{0}$, respectively, by the implicit function theorem (Newey and Smith, 2004, proof of Theorem 2.1).

\begin{lemma}
Suppose Assumptions 1-3 hold with $q_{1}\geq 2$ and $q_{1}>2a$ for some $a\geq 0$. Then, for all $\varepsilon>0$,
\begin{eqnarray}
\nonumber (a) && \lim_{n\rightarrow\infty}n^{a}P\left(\sup_{\theta\in\Theta}\left\|\hat{\lambda}(\theta)-\lambda_{0}(\theta)\right\|>\varepsilon\right)=0,\\
\nonumber (b) &&
\lim_{n\rightarrow\infty}n^{a}P\left(\sup_{\theta\in\Theta}\left|n^{-1}\sum_{i=1}^{n}\left(\rho(\hat{\lambda}(\theta)'g_{i}(\theta))-E\rho(\lambda_{0}(\theta)'g_{i}(\theta))\right)\right|>\varepsilon\right)=0.
\end{eqnarray}
\label{L2}
\end{lemma}
\vspace{-2em}
\noindent
\textit{Proof.}
The proof is similar to those of Lemma 2 of HH and Theorem 10 of S2007. First, we need to show
\begin{eqnarray}
\lim_{n\rightarrow\infty}n^{a}P\left(\sup_{\theta\in\Theta}\sup_{\lambda\in\Lambda(\theta)}\left|n^{-1}\sum_{i=1}^{n}\left(\rho(\lambda'g_{i}(\theta))-E\rho(\lambda'g_{i}(\theta))\right)\right|>\varepsilon\right)=0.
\label{UWLLN1}
\end{eqnarray}
This is proved by the proof of Lemma 2 of HH with $\rho(\lambda'g_{i}(\theta))$ in place of their $G(x,\theta)$, except that we use AL1(a) instead of Lemma 1 of HH. In particular, we apply AL1(a) with $c=0$ and $h(X_{i})=C_{\rho}(X_{i})-EC_{\rho}(X_{i})$ or $h(X_{i})=\rho(\lambda_{j}'g_{i}(\theta_{j}))-E\rho(\lambda_{j}'g_{i}(\theta_{j}))$ for any $\lambda_{j}\in\Lambda(\theta_{j})$ and any $\theta_{j}\in\Theta$. Since a zero vector is in $\Lambda(\theta)$, $\Theta$ and $\Lambda(\theta)$ are compacts, and $\rho(0)=0$, Assumption \ref{A2}(d) implies that $E|\rho(\lambda'g_{i}(\theta))|^{q_{1}}<\infty$ for all $\lambda\in\Lambda(\theta)$ and all $\theta\in\Theta$. Thus, the conditions for AL1(a) is satisfied by letting $p=q_{1}$ and Assumption 2(d). 

Next, we show
\begin{equation}
\lim_{n\rightarrow\infty}n^{a}P\left(\sup_{\theta\in\Theta}\left\|\bar{\lambda}(\theta)-\lambda_{0}(\theta)\right\|>\varepsilon\right)= 0,
\label{lambar1}
\end{equation}
where $\bar{\lambda}(\theta)=\argmax_{\lambda\in\Lambda(\theta)} n^{-1}\sum_{i=1}^{n}\rho(\lambda'g_{i}(\theta))$. This is proved by using Step 1 of the proof of Theorem 10 of S2007 and \eqref{UWLLN1}. Then, the present lemma (a) is proved by a similar argument with the proof of Theorem 2.7 of Newey and McFadden (1994) using the concavity of $n^{-1}\sum_{i=1}^{n}\rho(\lambda'g_{i}(\theta))$ in $\lambda$ for any $\theta$.

Finally, the present lemma (b) can be shown as follows. By the triangle inequality, combining the following results proves the desired result.
\begin{eqnarray}
&& \lim_{n\rightarrow\infty}n^{a}P\left(\sup_{\theta\in\Theta}\left|n^{-1}\sum_{i=1}^{n}\rho(\hat{\lambda}(\theta)'g_{i}(\theta))-n^{-1}\sum_{i=1}^{n}\rho(\lambda_{0}(\theta)'g_{i}(\theta))\right|>\varepsilon\right)=0,
\label{sub1}\\
&& \lim_{n\rightarrow\infty}n^{a}P\left(\sup_{\theta\in\Theta}\left|n^{-1}\sum_{i=1}^{n}\rho(\lambda_{0}(\theta)'g_{i}(\theta))-E\rho(\lambda_{0}(\theta)'g_{i}(\theta))\right|>\varepsilon\right)=0.
\label{sub2}
\end{eqnarray}
By Assumption 2(d), \eqref{sub1} follows from the present lemma (a) and AL1(b). Since $\lambda_{0}(\theta)\in int(\Lambda(\theta))$, \eqref{sub2} follows from \eqref{UWLLN1}.\qed
\vspace{1em}

Let $\tau_{0}\equiv Ee^{\lambda_{0}'g_{i}(\theta_{0})}$ and $\kappa_{0}\equiv-(Ee^{\lambda_{0}'g_{i}(\theta_{0})}g_{i}(\theta_{0})g_{i}(\theta_{0})')^{-1}\tau_{0}Eg_{i}(\theta_{0})$ for ETEL. Let $g$ and $G^{(j)}$ be an element of $g_{i}(\theta)$ and $G_{i}^{(j)}(\theta)$, respectively, for $j=1,...,d+1$. In addition, let $g^{k}$ be a multiplication of any $k$-combination of elements of $g_{i}(\theta)$. For instance, if $g_{i}(\theta)=(g_{i,1}(\theta),g_{i,2}(\theta))'$, a $2\times 1$ vector, then $g^{2}=(g_{i,1}(\theta))^{2}$, $g_{i,1}(\theta)g_{i,2}(\theta)$, or $(g_{i,2}(\theta))^{2}$. $G^{(j)k}$ is defined analogously. 

\begin{lemma}
Suppose Assumptions 1-3 hold with $q_{1}\geq 2$, $q_{1}>2a$, and $q_{2}>\max\left\{2,\frac{2a}{1-2c}\right\}$ for some $c\in[0,1/2)$ and some $a\geq 0$.
Then,
\begin{equation}
\nonumber \lim_{n\rightarrow\infty}n^{a}P\left(\|\hat{\beta}-\beta_{0}\|>n^{-c}\right)=0,
\end{equation}
where $\hat{\beta}=(\hat{\theta}',\hat{\lambda}')'$ and $\beta_{0}=(\theta_{0}',\lambda_{0}')'$ for EL and ET, and $\hat{\beta}=(\hat{\theta}',\hat{\lambda}',\hat{\kappa}',\hat{\tau})'$ and $\beta_{0}=(\theta_{0}',\lambda_{0}',\kappa_{0}',\tau_{0})'$ for ETEL.
\label{L3}
\end{lemma}
\noindent
\textit{Proof.}
We first show for any $\varepsilon>0$,
\begin{equation}
\lim_{n\rightarrow\infty}n^{a}P\left(\|\hat{\beta}-\beta_{0}\|>\varepsilon\right)=0.
\label{L3_ep}
\end{equation}
First, consider EL or ET. Since $\rho(\lambda_{0}(\theta)'g_{i}(\theta))$ is continuous in $\theta$ and uniquely minimized at $\theta_{0}\in int(\Theta)$, standard consistency arguments using Lemma \ref{L2}(b) show that 
\begin{equation}
\lim_{n\rightarrow\infty}n^{a}P\left(\|\hat{\theta}-\theta_{0}\|>\varepsilon\right)=0.
\label{con_theta}
\end{equation}
Write $\hat{\lambda}\equiv\hat{\lambda}(\hat{\theta})$ and $\lambda_{0}\equiv\lambda_{0}(\theta_{0})$. By Lemma \ref{L2}(a), \eqref{con_theta}, and the implicit function theorem that $\lambda_{0}(\theta)$ is continuous in a neighborhood of $\theta_{0}$, it follows
\begin{equation}
\lim_{n\rightarrow\infty}n^{a}P\left(\|\hat{\lambda}-\lambda_{0}\|>\varepsilon\right)=0.
\label{con_lambda}
\end{equation}
This proves \eqref{L3_ep} for EL and ET. For ETEL, \eqref{con_theta} and \eqref{con_lambda} can be shown by Step 2 of the proof of Theorem 10 of S2007 by applying AL1, AL2, and Lemma \ref{L2}.  Since we have introduced auxiliary parameters $(\kappa,\tau)$ for ETEL, we need to prove consistency of $(\hat{\kappa},\hat{\tau})$ as well. The proof is straightforward because they are continuous functions of $\hat{\lambda}$ and $\hat{g}_{i}$. We apply \eqref{con_theta}, \eqref{con_lambda}, AL1, the triangle inequality, the Schwarz matrix inequality, H\"{o}lder's inequality, and Assumptions \ref{A2}(c) and \ref{A3}(d). Note that Assumption \ref{A3}(d) implies $Ee^{q_{2}\lambda_{0}'g_{i}(\theta_{0})}<\infty$ for $q_{2}>\max\{2,2a\}$, because (i) a zero vector is in $\Lambda(\theta)$, (ii) $\Theta$ and $\Lambda(\theta)$ are compacts, and (iii) $\rho(0)=0$. 

Since we have established consistency of $\hat{\beta}$ for $\beta_{0}$, we now show the present lemma. The proof is similar to that of Lemma 3 of A2002 and Step 3 of the proof of Theorem 10 of S2007. Since $\hat{\beta}$ is in the interior of the compact sets with probability $1-o(n^{-a})$, $\hat{\beta}$ is the solution to $n^{-1}\sum_{i=1}^{n}\psi(X_{i},\hat{\beta})=0$ with probability $1-o(n^{-a})$. By the mean value expansion of $n^{-1}\sum_{i=1}^{n}\psi(X_{i},\hat{\beta})=0$ around $\beta_{0}$,
\begin{equation}
\hat{\beta}-\beta_{0} = -\left(n^{-1}\sum_{i=1}^{n}\frac{\partial\psi(X_{i},\tilde{\beta})}{\partial\beta'}\right)^{-1}n^{-1}\sum_{i=1}^{n}\psi(X_{i},\beta_{0}),
\end{equation}
with probability $1-o(n^{-a})$, where $\tilde{\beta}$ lies between $\hat{\beta}$ and $\beta_{0}$ and may differ across rows. The lemma follows from
\begin{eqnarray}
&& \lim_{n\rightarrow\infty}n^{a}P\left(\left\|n^{-1}\sum_{i=1}^{n}\frac{\partial\psi(X_{i},\tilde{\beta})}{\partial\beta'}-n^{-1}\sum_{i=1}^{n}\frac{\partial\psi(X_{i},\beta_{0})}{\partial\beta'}\right\|>\varepsilon\right)=0,
\label{NC-1}\\
&& \lim_{n\rightarrow\infty}n^{a}P\left(\left\|n^{-1}\sum_{i=1}^{n}\frac{\partial\psi(X_{i},\beta_{0})}{\partial\beta'}-E\frac{\partial\psi(X_{i},\beta_{0})}{\partial\beta'}\right\|>\varepsilon\right)=0,
\label{NC-2}\\
&& \lim_{n\rightarrow\infty}n^{a}P\left(\left\|n^{-1}\sum_{i=1}^{n}\psi(X_{i},\beta_{0})\right\|>n^{-c}\right)=0.
\label{NC-3}
\end{eqnarray}
First, to show \eqref{NC-1}, observe that the elements of $(\partial/\partial\beta')\psi(X_{i},\beta)$ have the form
\begin{equation}
\alpha\cdot\rho_{j}^{k_{\rho}}(\lambda'g_{i})\cdot g^{k_{0}}\cdot G^{k_{1}}\cdot G^{(2)k_{2}},\hspace{0.5em}j=1,2,
\end{equation}
where $\alpha$ denotes products of components of $\beta$, $k_{\rho}=1$, $k_{0}\leq 2$, $k_{1}\leq 2$, and $k_{2}\leq 1$ for EL and ET. For ETEL, we replace $\rho_{j}^{k_{\rho}}(\lambda'g_{i0})$ with $e^{k_{\rho}\lambda_{0}'g_{i0}}$, where $k_{\rho}=0,1$, $k_{0}\leq 3$, $k_{1}\leq 2$, and $k_{2}\leq 1$. For each element, we apply the triangle inequality, \eqref{L3_ep}, and AL1(b) multiple times. The condition of AL1(b) is satisfied by Assumptions 2-3, H\"{o}lder's inequality, and letting $p=q_{2}$. This proves \eqref{NC-1}.  The second result \eqref{NC-2} can be shown analogously by using AL1(a) with $c=0$ and $h(X_{i}) = (\partial/\partial\beta')\psi(X_{i},\beta_{0})- E(\partial/\partial\beta')\psi(X_{i},\beta_{0})$. The last result \eqref{NC-3} holds by AL1(a) with $h(X_{i})=\psi(X_{i},\beta_{0})$. For example, $e^{\lambda_{0}'g_{i0}}g_{i0}$ is an element of $\psi(X_{i},\beta_{0})$, and it needs to satisfy the condition of AL1(a) with $h(X_{i})=e^{\lambda_{0}'g_{i0}}g_{i0}$. By using H\"{o}lder's inequality, 
\begin{equation}
Ee^{p\cdot \lambda_{0}'g_{i0}}(X_{i})\|g_{i0}\|^{p}\leq \left(Ee^{p(1+\epsilon)\lambda_{0}'g_{i0}}\right)^{\frac{1}{1+\epsilon}}\cdot\left(E\|g_{i0}\|^{p(1+\epsilon^{-1})}\right)^{\frac{\epsilon}{1+\epsilon}},
\label{holder11}
\end{equation}
for any $0<\epsilon<\infty$. Since Assumption \ref{A2}(c) holds for all $0<q_{g}<\infty$, given $a$ and $c$, we can take small enough $\epsilon$ so that $p=q_{2}>\max\{2,\frac{2a}{1-2c}\}$ implies that \eqref{holder11} is finite by Assumption \ref{A3}(d). Other elements of $\psi(X_{i},\beta_{0})$ can be shown similarly.\qed

\vspace{1em}

Write $g_{i}^{*}(\theta)\equiv g(X_{i}^{*},\theta)$ and $\hat{g}^{*}_{i}\equiv g^{*}(\hat{\theta}^{*})$. Define $\hat{\lambda}^{*}(\theta)=\argmax_{\lambda\in\mathbf{R}^{L_{g}}} n^{-1}\sum_{i=1}^{n}\rho(\lambda'g_{i}^{*}(\theta))$ and write $\hat{\lambda}^{*}\equiv\hat{\lambda}^{*}(\hat{\theta}^{*})$. Let $\rho(\nu)=1-e^{\nu}$ for ETEL in the next lemma.

\begin{lemma}
Suppose Assumptions 1-3 hold with $q_{1}\geq 2$ and $q_{1}>4a$ for some $a\geq0$. Then, for all $\varepsilon>0$,
\begin{eqnarray}
\nonumber (a) && \lim_{n\rightarrow\infty}n^{a}P\left(P^{*}\left(\sup_{\theta\in\Theta}\left\|\hat{\lambda}^{*}(\theta)-\hat{\lambda}(\theta)\right\|>\varepsilon\right)>n^{-a}\right)=0,\\
\nonumber (b) && \lim_{n\rightarrow\infty}n^{a}P\left(P^{*}\left(\sup_{\theta\in\Theta}\left|n^{-1}\sum_{i=1}^{n}\left(\rho(\hat{\lambda}^{*}(\theta)'g_{i}^{*}(\theta))-\rho(\hat{\lambda}(\theta)'g_{i}(\theta))\right)\right|>\varepsilon\right)>n^{-a}\right)=0.
\end{eqnarray}
\label{L4}
\end{lemma}
\vspace{-2em}
\noindent
\textit{Proof.}
We first show
\begin{equation}
\lim_{n\rightarrow\infty}n^{a}P\left(P^{*}\left(\sup_{\theta\in\Theta}\sup_{\lambda\in\Lambda(\theta)}\left|n^{-1}\sum_{i=1}^{n}\left(\rho(\lambda'g_{i}^{*}(\theta))-\rho(\lambda'g_{i}(\theta))\right)\right|>\varepsilon\right)>n^{-a}\right)=0.
\label{bootuni1}
\end{equation}
We use the proof of Lemma 8 of HH using AL6(a) with $c=0$. Since $n^{-1}\sum_{i=1}^{n}\rho(\lambda'g_{i}(\theta))=E^{*}\rho(\lambda'g_{i}^{*}(\theta))$, we apply AL6(a) with $h(X_{i})=\rho(\lambda_{j}'g_{i}(\theta_{j}))-E\rho(\lambda_{j}'g_{i}(\theta_{j}))$ for any $\lambda_{j}\in\Lambda(\theta_{j})$ and $\theta_{j}\in\Theta$ or $h(X_{i})=C_{\rho}(X_{i})-EC_{\rho}(X_{i})$. By Minkowski inequality, it suffices to show $E|\rho(\lambda_{j}'g_{i}(\theta_{j}))|^{p}<\infty$ and $EC_{\rho}^{p}(X_{i})<\infty$ for $p\geq 2$ and $p>4a$. This holds by letting $p=q_{1}$.

Next, we show
\begin{equation}
\lim_{n\rightarrow\infty}n^{a}P\left(P^{*}\left(\sup_{\theta\in\Theta}\left\|\bar{\lambda}^{*}(\theta)-\bar{\lambda}(\theta)\right\|>\varepsilon\right)>n^{-a}\right)=0,
\label{Un2}
\end{equation}
where $\bar{\lambda}^{*}(\theta)=\argmax_{\lambda\in\Lambda(\theta)}n^{-1}\sum_{i=1}^{n}\rho(\lambda'g_{i}^{*}(\theta))$. We claim that for a given $\varepsilon>0$, there exists $\eta>0$ independent of $n$ such that for any $\theta\in\Theta$ and any $\lambda\in\Lambda(\theta)$, $\|\lambda-\bar{\lambda}(\theta)\|>\varepsilon$ implies that $n^{-1}\sum_{i}\rho(\bar{\lambda}(\theta)'g_{i}(\theta))-n^{-1}\sum_{i}\rho(\lambda'g_{i}(\theta)) \geq \eta >0$ with probability $1-o(n^{-a})$. This claim can be shown by similar arguments with the proof of Lemma 9 of A2002. For any $\theta\in\Theta$ and any $\lambda\in\Lambda(\theta)$, whenever $\|\lambda-\bar{\lambda}(\theta)\|>\varepsilon$, $\|\lambda-\lambda_{0}(\theta)\|>\varepsilon/2$ with probability $1-o(n^{-a})$ by the triangle inequality and Lemma \ref{L2}. Since, for a given $\theta$, $E\rho(\lambda'g_{i}(\theta))$ is uniquely maximized at $\lambda_{0}(\theta)$ and continuous on $\Lambda(\theta)$, $\|\lambda-\lambda_{0}(\theta)\|>\varepsilon/2$ implies that there exists $\eta(\theta)>0$ such that
\begin{eqnarray}
\eta(\theta)&\leq& E\rho(\lambda_{0}(\theta)'g_{i}(\theta))-E\rho(\lambda'g_{i}(\theta))
\label{eta_ind}\\
\nonumber &\leq& n^{-1}\sum_{i}\left(\rho(\bar{\lambda}(\theta)'g_{i}(\theta))-\rho(\lambda'g_{i}(\theta))\right)+ 2\sup_{\theta\in\Theta}\sup_{\lambda\in\Lambda(\theta)}|n^{-1}\sum_{i}\rho(\lambda'g_{i}(\theta))-E\rho(\lambda'g_{i}(\theta))|.
\end{eqnarray}
Since \eqref{UWLLN1} holds for all $\varepsilon$ and $\Theta$ is a compact set, letting $\varepsilon=\eta(\theta)/3$ in \eqref{UWLLN1} and $\eta=\inf_{\theta}\eta(\theta)$ proves the claim. Then, we have
\begin{eqnarray}
&& P(P^{*}(\sup_{\theta\in\Theta}\|\bar{\lambda}^{*}(\theta)-\bar{\lambda}(\theta)\|>\varepsilon)>n^{-a})\\
\nonumber &\leq& P\left(P^{*}\left(\sup_{\theta\in\Theta}\left|n^{-1}\sum_{i}\left(\rho(\bar{\lambda}(\theta)'g_{i}(\theta))-\rho(\bar{\lambda}^{*}(\theta)'g_{i}(\theta))\right)\right|>\eta\right)>n^{-a}\right)\\
\nonumber &\leq& P\left(P^{*}\left(\sup_{\theta\in\Theta}\sup_{\lambda\in\Lambda(\theta)}\left|n^{-1}\sum_{i}\left(\rho(\lambda'g_{i}^{*}(\theta))-\rho(\lambda'g_{i}(\theta))\right)\right|>\eta/2\right)>n^{-a}\right) = o(n^{-a}).
\end{eqnarray}
The second inequality holds by adding and subtracting $n^{-1}\sum_{i}\rho(\bar{\lambda}(\theta)'g_{i}^{*}(\theta))$, and using the definition of $\bar{\lambda}^{*}(\theta)$. The last equality follows by \eqref{bootuni1}. The present lemma (a) can be obtained by replacing $\bar{\lambda}^{*}(\theta)$ and $\bar{\lambda}(\theta)$ with $\hat{\lambda}^{*}(\theta)$ and $\hat{\lambda}(\theta)$, respectively. Since $n^{-1}\sum_{i}\rho(\lambda'g_{i}(\theta))$ and $n^{-1}\sum_{i}\rho(\lambda'g_{i}^{*}(\theta))$ are concave in $\lambda$ for any $\theta$, as long as $\bar{\lambda}(\theta)$ and $\bar{\lambda}^{*}(\theta)$ are in the interior of $\Lambda(\theta)$, they are maximizers on $\mathbf{R}^{L_{g}}$ by Theorem 2.7 of Newey and McFadden (1994). But by Assumption 2, $\bar{\lambda}(\theta)\in int(\Lambda(\theta))$ with probability $1-o(n^{-a})$ and $\bar{\lambda}^{*}(\theta)\in int(\Lambda(\theta))$ with $P^{*}$ probability $1-o(n^{-a})$ except, possibly, if $\chi_{n}$ is in a set of  $P$ probability $o(n^{-a})$. Therefore, the present lemma (a) is proved.
 
Finally, the present Lemma (b) follows from the results below:
\begin{eqnarray}
 \lim_{n\rightarrow\infty}n^{a}P\left(P^{*}\left(\sup_{\theta\in\Theta}\left|n^{-1}\sum_{i=1}^{n}\left(\rho(\hat{\lambda}^{*}(\theta)'g_{i}^{*}(\theta))-\rho(\hat{\lambda}(\theta)'g_{i}^{*}(\theta))\right)\right|>\varepsilon\right)>n^{-a}\right)=0,
\label{Bsub1}\\
 \lim_{n\rightarrow\infty}n^{a}P\left(P^{*}\left(\sup_{\theta\in\Theta}\left|n^{-1}\sum_{i=1}^{n}\left(\rho(\hat{\lambda}(\theta)'g_{i}^{*}(\theta))-\rho(\hat{\lambda}(\theta)'g_{i}(\theta))\right)\right|>\varepsilon\right)>n^{-a}\right)=0.
\label{Bsub2}
\end{eqnarray}
\eqref{Bsub1} can be shown as follows. By Assumption 2(d) and standard manipulation,
\begin{eqnarray}
&& P\left(P^{*}\left(\sup_{\theta\in\Theta}\left|n^{-1}\sum_{i=1}^{n}\left(\rho(\hat{\lambda}^{*}(\theta)'g_{i}^{*}(\theta))-\rho(\hat{\lambda}(\theta)'g_{i}^{*}(\theta))\right)\right|>\varepsilon\right)>n^{-a}\right)\\
\nonumber 
&\leq& P\left(P^{*}\left(n^{-1}\sum_{i}C_{\rho}(X_{i}^{*})>\varepsilon\right)>n^{-a}/2\right) + P\left(P^{*}\left(\sup_{\theta\in\Theta}\|\hat{\lambda}^{*}(\theta)-\hat{\lambda}(\theta)\|>1\right)>n^{-a}/2\right).
\end{eqnarray}
We apply AL6(d) with $h(X_{i})=C_{\rho}(X_{i})$ and $p=q_{1}$ for the first term on the RHS of the above inequality, and apply the present lemma (a) for the second term to show that the RHS is $o(n^{-a})$. This proves \eqref{Bsub1}. Since $\hat{\lambda}(\theta)\in int(\Lambda(\theta))$ with probability $1-o(n^{-a})$, \eqref{Bsub2} follows from \eqref{bootuni1}.\qed

\begin{lemma}
Suppose Assumptions 1-3 hold with $q_{1}\geq 2$, $q_{1}>4a$, and $q_{2}>\max\left\{2,\frac{4a}{1-2c}\right\}$ for some $c\in[0,1/2)$ and some $a\geq 0$. Then,
\begin{equation}
\nonumber \lim_{n\rightarrow\infty}n^{a}P\left(P^{*}\left(\|\hat{\beta}^{*}-\hat{\beta}\|>n^{-c}\right)>n^{-a}\right)=0,
\end{equation}
where $\hat{\beta}^{*}=(\hat{\theta}^{*'},\hat{\lambda}^{*'})'$ and $\hat{\beta}=(\hat{\theta}',\hat{\lambda}')'$ for EL and ET, and $\hat{\beta}^{*}=(\hat{\theta}^{*'},\hat{\lambda}^{*'},\hat{\kappa}^{*'},\hat{\tau}^{*})'$ and $\hat{\beta}=(\hat{\theta}',\hat{\lambda}',\hat{\kappa}',\hat{\tau})'$ for ETEL.
\label{L5}
\end{lemma}
\noindent
\textit{Proof.}
The proof is analogous to that of Lemma \ref{L3} except that it involves additional steps for the bootstrap versions of the estimators. First, we show
\begin{equation}
\lim_{n\rightarrow\infty}n^{a}P\left(P^{*}\left(\|\hat{\beta}^{*}-\hat{\beta}\|>\varepsilon\right)>n^{-a}\right)=0.
\label{con_boot1}
\end{equation}
Consider EL or ET. We claim that for a given $\varepsilon>0$, there exists $\eta>0$ independent of $n$ such that $\|\theta-\hat{\theta}\|>\varepsilon$ implies that $0<\eta\leq n^{-1}\sum_{i}\rho(\hat{\lambda}(\theta)'g_{i}(\theta))-n^{-1}\sum_{i}\rho(\hat{\lambda}'\hat{g}_{i})$ with probability $1-o(n^{-a})$. This claim can be shown by a similar argument with \eqref{eta_ind} by using the fact that $E\rho(\lambda_{0}(\theta)'g_{i}(\theta))$ is uniquely minimized at $\theta_{0}$ and continuous in $\theta$, AL1(b), Lemma \ref{L2}(a), \eqref{sub2}, \eqref{con_theta}, and \eqref{con_lambda}. Thus, we have
\begin{eqnarray}
&& P\left(P^{*}\left(\|\hat{\theta}^{*}-\hat{\theta}\|>\varepsilon\right)>n^{-a}\right)
\label{con_theboot1}
\\
\nonumber &\leq& P\left(P^{*}\left(\sup_{\theta\in\Theta}\left|n^{-1}\sum_{i=1}^{n}\left(\rho(\hat{\lambda}^{*}(\theta)'g_{i}^{*}(\theta))-\rho(\hat{\lambda}(\theta)'g_{i}(\theta))\right)\right|>\eta/2\right)>n^{-a}\right)=o(n^{-a}),
\end{eqnarray}
by Lemma \ref{L4}(b). To show
\begin{equation}
\lim_{n\rightarrow\infty}n^{a}P\left(P^{*}\left(\|\hat{\lambda}^{*}-\hat{\lambda}\|>\varepsilon\right)>n^{-a}\right)=0,
\label{con_lamboot1}
\end{equation}
we use the triangle inequality, \eqref{con_theta}, \eqref{con_theboot1}, Lemma \ref{L2}(a), Lemma \ref{L4}(a), and the implicit function theorem that $\lambda_{0}(\theta)$ is continuously differentiable around $\theta_{0}$. This proves \eqref{con_boot1} for EL or ET. For ETEL, an analogous result to Lemma \ref{L4}(b), 
\begin{equation}
\lim_{n\rightarrow\infty}n^{a}P\left(P^{*}\left(\sup_{\theta\in\Theta}\left|n^{-1}\sum_{i=1}^{n}\left(\hat{l}^{*}_{n}(\theta)-\hat{l}_{n}(\theta)\right)\right|>\varepsilon\right)>n^{-a}\right)=0,
\end{equation}
where
\begin{equation}
\hat{l}_{n}^{*}(\theta) = \log \left(n^{-1}\sum_{i=1}^{n}e^{\hat{\lambda}^{*}(\theta)'(g_{i}^{*}(\theta)-\bar{g}_{n}^{*}(\theta))}\right),
\label{bootETEL1}
\end{equation}
and $\bar{g}_{n}^{*}(\theta)=n^{-1}\sum_{i=1}^{n}g_{i}^{*}(\theta)$, can be shown by Lemma \ref{L4}(a), AL6, and AL7. Then, replacing $E\rho(\lambda_{0}(\theta)'g_{i}(\theta))$ with $l_{0}(\theta)$ and $n^{-1}\sum_{i}\rho(\hat{\lambda}(\theta)'g_{i}(\theta))$ with $\hat{l}_{n}(\theta)$, and applying a similar argument with \eqref{eta_ind} give \eqref{con_theboot1} and \eqref{con_lamboot1} for ETEL. For the auxiliary parameters $\kappa$ and $\tau$, the bootstrap versions of the estimators are $\hat{\kappa}^{*} =  -(n^{-1}\sum_{i=1}^{n}e^{\hat{\lambda}^{*'}\hat{g}^{*}_{i}}\hat{g}^{*}_{i}\hat{g}^{*'}_{i})^{-1}\hat{\tau}^{*}\hat{\bar{g}}^{*}_{n}$ and
$\hat{\tau}^{*} = n^{-1}\sum_{i=1}^{n}e^{\hat{\lambda}^{*'}\hat{g}^{*}_{i}}$, where $\hat{\bar{g}}^{*}_{n}=n^{-1}\sum_{i=1}^{n}\hat{g}_{i}$. Since they are continuous functions of $\hat{\lambda}^{*}$ and $\hat{g}_{i}^{*}$, analogous results to \eqref{con_theboot1} and \eqref{con_lamboot1} can be shown by the triangle inequality, AL6-AL7, Lemma \ref{L4}, and the implicit function theorem that $\hat{\lambda}^{*}(\theta)$ is continuously differentiable around $\hat{\theta}^{*}$.

The rest of the proof to show the argument of the lemma (with $n^{-c}$ in place of $\varepsilon$) is analogous to that of Lemma \ref{L3} except that we apply AL6 instead of AL1. By H\"{o}lder's inequality, the binding condition is $p=q_{2}>\max\left\{2,4a/(1-2c)\right\}$ for AL6 but this is satisfied by the assumption of the lemma. \qed

\vspace{1em}
Let $S_{n}$ be a vector containing the unique components of $n^{-1}\sum_{i=1}^{n}f(X_{i},\beta_{0})$ on the support of $X_{i}$, and $S=ES_{n}$. Similarly, let $S^{*}_{n}$ denote a vector containing the unique components of $n^{-1}\sum_{i=1}^{n}f(X_{i}^{*},\hat{\beta})$ on the support of $X_{i}$, and $S^{*}=E^{*}S_{n}^{*}$.

\begin{lemma}
(a) Suppose Assumptions 1-3 hold with $q_{2}>\max\left\{4,4a\right\}$ for some $a\geq 0$. Then, for all $\varepsilon>0$,
\begin{equation}
\nonumber \lim_{n\rightarrow\infty}n^{a}P\left(\|S_{n}-S\|>\varepsilon\right)=0.
\end{equation}
(b) Suppose Assumptions 1-3 hold with $q_{1}\geq 2$, $q_{1}>2a$, and $q_{2}>\max\left\{4,8a\right\}$ for some $a\geq 0$. Then, for all $\varepsilon>0$,
\begin{equation}
\nonumber \lim_{n\rightarrow\infty}n^{a}P\left(P^{*}\left(\|S_{n}^{*}-S^{*}\|>\varepsilon\right)>n^{-a}\right)=0.
\end{equation}
\label{L6}
\end{lemma}
\vspace{-1em}
\noindent
\textit{Proof.} The present lemma (a) can be shown as follows. Let $s_{i}(\beta_{0})$ be the least favorable term in $f(X_{i},\beta_{0})$ with respect to the value of $q_{2}$. Write $s_{i}\equiv s_{i}(\beta_{0})$. Then it suffices to show
\begin{equation}
P\left(\left\|n^{-1}\sum_{i=1}^{n}s_{i}-Es_{i}\right\|>\varepsilon\right)=o(n^{-a}).
\label{f1}
\end{equation}
We apply AL1(b) with $c=0$ and $h(X_{i})=s_{i}-Es_{i}$. To see what $s_{i}$ is, we need to spell out the components of $f(X_{i},\beta_{0})$. For EL or ET, $f(X_{i},\beta)$ consists of terms of the form 
\begin{equation}
\alpha\cdot \rho_{j}^{k_{\rho}}(\lambda'g_{i}(\theta))\cdot g^{k_{0}}\cdot G^{k_{1}}\cdots G^{(d+1)k_{d+1}},
\label{term1}
\end{equation}
where $\alpha$ denotes products of components of $\beta$ and and $k_{l}$'s are nonnegative integers for $l=0,1,...d+1$. In addition, $j=1,...,d+1$, $k_{\rho}=1,2$, $k_{0},k_{1}\leq d+1$, $k_{l}\leq d-l+1$ for $l=2,...,d$, $k_{d+1}\leq 1$, and $\sum_{l=0}^{d+1}k_{l}\leq d+1$. For ETEL, we replace $\rho_{j}^{k_{\rho}}(\lambda'g_{i}(\theta))$ with $e^{k_{\rho}\lambda'g_{i}(\theta)}$, where $k_{\rho}=0,1,2$, $k_{0}\leq d+3$, $k_{l}\leq d-l+2$ for $l=1,2,...,d+1$, and $\sum_{l=0}^{d+1}k_{l}\leq d+3$. Since we assume that all the moments are finite for $g_{i}(\theta)$, $\forall\theta\in\Theta$ and $G_{i0}^{(j)}$, $j=1,2,...,d+1$, the values of $k_{l}$'s do not impose additional restriction on $q_{g}$ and $q_{G}$. What matters is $k_{\rho}$, which is directly related to $q_{2}$ in Assumption \ref{A3}(d). Since $k_{\rho}=2$ is the most restrictive case, it suffices to show $EC_{\partial\rho}^{2p}(X_{i})C_{g}^{(d+3)p}(X_{i})<\infty$, $EC_{\partial\rho}^{2p}(X_{i})C_{G}^{(d+3)p}(X_{i})<\infty$, $Ee^{2p\lambda_{0}'g_{i0}}C_{g}^{(d+3)p}(X_{i})<\infty$ and $Ee^{2p\lambda_{0}'g_{i0}}C_{G}^{(d+3)p}(X_{i})<\infty$ for AL1(b) to be applied. By H\"{o}lder's inequality, letting $p=q_{2}>\max\left\{4,4a\right\}$ satisfies these conditions. 

The present lemma (b) can be shown as follows. Let $s_{i}^{*}(\beta)$ be the least favorable term in $f(X_{i}^{*},\beta)$ with respect to the value of $q_{2}$ and write $\hat{s}_{i}^{*}\equiv s_{i}^{*}(\hat{\beta})$, $s_{i}^{*}\equiv s_{i}^{*}(\beta_{0})$, and $\hat{s}_{i}\equiv s_{i}(\hat{\beta})$. It suffices to show
\begin{equation}
P\left(P^{*}\left(\left\|n^{-1}\sum_{i=1}^{n}\hat{s}_{i}^{*}-n^{-1}\sum_{i=1}^{n}\hat{s}_{i}\right\|>\varepsilon\right)>n^{-a}\right)=o(n^{-a}).
\label{resulb}
\end{equation}
By the triangle inequality,
\begin{equation}
\left\|n^{-1}\sum_{i=1}^{n}\left(\hat{s}_{i}^{*}-\hat{s}_{i}\right)\right\| \leq \left\|n^{-1}\sum_{i=1}^{n}\left(s_{i}^{*}-s_{i}\right)\right\|+ \left\|n^{-1}\sum_{i=1}^{n}\left(\hat{s}_{i}^{*}-s_{i}^{*}\right)\right\| + \left\|n^{-1}\sum_{i=1}^{n}\left(\hat{s}_{i}-s_{i}\right)\right\|. \label{lem101}
\end{equation}
For the first term of the RHS of \eqref{lem101}, we apply Lemma AL6(a) with $c=0$ and $h(X_{i})=s_{i}-Es_{i}$. By using a similar argument with the proof of \eqref{f1}, the most restrictive condition is met with $p=q_{2}>\max\left\{4,8a\right\}$. The second and the last terms are shown by combining Lemma \ref{L3} with $c=0$ and the following results: For all $\beta\in N(\beta_{0})$, some neighborhood of $\beta_{0}$, there exist some functions $C(X_{i})$ and $C^{*}(X_{i}^{*})$ such that 
$\left\|s_{i}(\beta)-s_{i}\right\|\leq C(X_{i})\|\beta-\beta_{0}\|$ and $\left\|s_{i}^{*}(\beta)-s_{i}^{*}\right\|\leq C^{*}(X_{i}^{*})\|\beta-\beta_{0}\|$ and these functions satisfy for some $K<\infty$, $P(\|n^{-1}\sum_{i=1}^{n}C(X_{i})\|>K)=o(n^{-a})$ and $P(P^{*}(\|n^{-1}\sum_{i=1}^{n}C^{*}(X_{i}^{*})\|>K)>n^{-a})=o(n^{-a}).$ After some tedious but straightforward calculation using the binomial theorem, the triangle inequality, and H\"{o}lder's inequality, AL1(b) implies that the most restrictive case for the existence of such $C(X_{i})$ occurs when $k_{\rho}=2$, which is satisfied with $p=q_{2}>\max\left\{4,4a\right\}$. Similarly, the condition of AL6(d) with $h(X_{i}^{*})=C^{*}(X_{i}^{*})$ is satisfied with $p=q_{2}>\max\left\{4,8a\right\}$. \qed

\vspace{1em}

Lemma \ref{L7} shows the sample and the bootstrap versions of $t$ and Wald statistics are well approximated by smooth functions. Let $H_{n}(\theta)=((\partial/\partial\theta')\eta(\theta)\hat{\Sigma}_{MR}((\partial/\partial\theta')\eta(\theta))')^{-1/2}n^{1/2}\eta(\theta)$ and $H_{n}^{*}(\theta)=((\partial/\partial\theta')\eta(\theta)\hat{\Sigma}_{MR}^{*}((\partial/\partial\theta')\eta(\theta))')^{-1/2}n^{1/2}(\eta(\theta)-\eta(\hat{\theta}))$ so that $\mathcal{W}_{MR}=H_{n}(\hat{\theta})'H_{n}(\hat{\theta})$ and $\mathcal{W}_{MR}^{*}=H_{n}^{*}(\hat{\theta}^{*})'H_{n}^{*}(\hat{\theta}^{*})$.

\begin{lemma}
Let $\Delta_{n}$ and $\Delta_{n}^{*}$ denote $\sqrt{n}(\hat{\theta}-\theta_{0})$ and $\sqrt{n}(\hat{\theta}^{*}-\hat{\theta})$, or $T_{MR}$ and $T_{MR}^{*}$, or $H_{n}(\hat{\theta})$ and $H_{n}^{*}(\hat{\theta}^{*})$. For each definition of $\Delta_{n}$ and $\Delta_{n}^{*}$, there is an infinitely differentiable function $A(\cdot)$ with $A(S)=0$ and $A(S^{*})=0$ such that the following results hold.\\
(a) Suppose Assumptions 1-4 hold with $q_{1}\geq 2$, $q_{1}>2a$, and $q_{2}>\max\left\{4,4a,\frac{2ad}{d-2a-1}\right\}$ and $d\geq 2a+2$ for some $a\geq 0$, where $2a$ is a positive integer. Then,
\[\lim_{n\rightarrow\infty}\sup_{z}n^{a}|P(\Delta_{n}\leq z)-P(\sqrt{n}A(S_{n})\leq z)|=0.\]
(b) Suppose Assumptions 1-4 hold with $q_{1}\geq 2$, $q_{1}>4a$, and $q_{2}>\max\left\{4,8a,\frac{4ad}{d-2a-1}\right\}$ and $d\geq 2a+2$ for some $a\geq 0$, where $2a$ is a positive integer. Then,
\[\lim_{n\rightarrow\infty}n^{a}P\left(\sup_{z}|P^{*}(\Delta_{n}^{*}\leq z)-P^{*}(\sqrt{n}A(S_{n}^{*})\leq z)|>n^{-a}\right)=0.\]
\label{L7}
\end{lemma}
\vspace{-2em}
\noindent
\textit{Proof.} The proof is analogous to that of Lemma 13(a) of A2002 that uses his Lemmas 1 and 3-9. His Lemmas 1, 5, 6, and 7 are used in the proof, and denoted by AL1, AL5, AL6, and AL7, respectively. His Lemma 3 is replaced by our Lemma \ref{L3}. His Lemmas 4 and 8 are not required because GEL is a one-step estimator without a weight matrix. His Lemma 9 is replaced by our Lemma \ref{L5}. The main difference is that the conditions on $q_{1}$ and $q_{2}$ do not appear in the proof of A2002 for GMM. Lemma \ref{L6} is used to give conditions for $q_{1}$ and $q_{2}$. I provide a sketch of the proof and an explanation where the conditions of the lemma are derived from. 

For part (a), the proof proceeds by taking Taylor expansion of the FOC around $\beta_{0}$ through order $d-1$. The remainder term $\zeta_{n}$ from the Taylor expansion satisfies $\|\zeta_{n}\|\leq M\|\hat{\beta}-\beta_{0}\|^{d}\leq n^{-dc}$ for some $M<\infty$ with probability $1-o(n^{-a})$ by Lemma \ref{L3}. To apply AL5(a), the conditions such that $n^{-dc+1/2}=o(n^{-a})$ or $dc\geq a+1/2$ for some $c\in[0,1/2)$, and that $2a$ is an integer, need to be satisfied. The former is satisfied if $d> 2a+1$ or $d\geq 2a+2$ (both $d$ and $2a$ are integers), and the latter is assumed. The condition on $q_{2}$ of Lemma \ref{L3} is minimized with the smallest $c$, let $c=(a+1/2)d^{-1}$. By plugging this into the condition of Lemma \ref{L3}, we have $q_{1}\geq 2$, $q_{1}>2a$, and $q_{2}>\max\{2,2ad(d-2a-1)^{-1}\}$. In addition, we use Lemma \ref{L6}(a) to use the implicit function theorem for the existence of $A(\cdot)$. By collecting the conditions of Lemmas \ref{L3} and \ref{L6}(a), we have the condition for the present lemma. The present lemma (a) for $\Delta_{n}=T_{MR}$ and $\Delta_{n}=H_{n}(\hat{\theta})$ can be shown similarly by using the fact that $\hat{\Sigma}_{MR}=\hat{\Sigma}_{MR}(\hat{\beta})$ is a function of $\hat{\beta}$.

The proof of part (b) proceeds analogously. By plugging the same $c$ into the condition of Lemma \ref{L5}, we have $q_{2}>\max\{2,4ad(d-2a-1)^{-1}\}$. The condition of Lemma \ref{L6}(b) is $q_{1}\geq 2$, $q_{1}>2a$, and $q_{2}>\max\{4,8a\}$. The condition of the present lemma collects these conditions. \qed

\vspace{1em}

We define the components of the Edgeworth expansions of the test statistics $T_{MR}$ and $\mathcal{W}_{MR}$ and their bootstrap analog $T_{MR}^{*}$ and $\mathcal{W}_{MR}^{*}$. Let $\Psi_{n}=\sqrt{n}(S_{n}-S)$ and $\Psi_{n}^{*}=\sqrt{n}(S_{n}^{*}-S^{*})$. Let $\Psi_{n,j}$ and $\Psi_{n,j}^{*}$ denote the $j$th elements of $\Psi_{n}$ and $\Psi_{n}^{*}$ respectively. Let $\nu_{n,a}$ and $\nu_{n,a}^{*}$ denote vectors of moments of the form $n^{\alpha(m)}E\Pi_{\mu=1}^{m}\Psi_{n,j_{\mu}}$ and $n^{\alpha(m)}E^{*}\Pi_{\mu=1}^{m}\Psi^{*}_{n,j_{\mu}}$, respectively, where $2\leq m\leq 2a+2$, $\alpha(m)=0$ if $m$ is even, and $\alpha(m)=1/2$ if $m$ is odd. Let $\nu_{a}=\lim_{n\rightarrow\infty}\nu_{n,a}$. The existence of the limit is proved in Lemma \ref{L8}.

Let $\pi_{i}(\delta,\nu_{a})$ be a polynomial in $\delta=\partial/\partial z$ whose coefficients are polynomials in the elements of $\nu_{a}$ and for which $\pi_{i}(\delta,\nu_{a})\Phi(z)$ is an even function of $z$ when $i$ is odd and is an odd function of $z$ when $i$ is even for $i=1,...,2a$, where $2a$ is an integer. The Edgeworth expansions of $T_{MR}$ and $T_{MR}^{*}$ depend on $\pi_{i}(\delta,\nu_{a})$ and $\pi_{i}(\delta,\nu_{n,a}^{*})$, respectively. In contrast, the Edgeworth expansions of $\mathcal{W}_{MR}$ and $\mathcal{W}_{MR}^{*}$ depend on $\pi_{\mathcal{W},i}(y,\nu_{a})$ and $\pi_{\mathcal{W},i}(y,\nu_{n,a}^{*})$ where $\pi_{\mathcal{W},i}(y,\nu_{a})$ is a polynomial in $y$ whose coefficients are polynomials in the elements of $\nu_{a}$ for $i=1,...,[a]$, and $[a]$ denotes the largest integer less than or equal to $a$. The following lemma provides conditions under which the bootstrap moments are close enough to the population moments in large samples.

\begin{lemma}
(a) Suppose Assumptions 1-3 hold with $q_{2}>4(a+1)$ for some $a\geq 0$. Then, $\nu_{n,a}$ and $\nu_{a}\equiv\lim_{n\rightarrow\infty}\nu_{n,a}$ exist.\\
(b) Suppose Assumptions 1-3 hold with $q_{1}\geq 2$, $q_{1}>2a$, and $q_{2}>\max\left\{8(a+1),\frac{8a(a+1)}{1-2\xi}\right\}$ for some $a\geq 0$ and some $\xi\in[0,1/2)$. Then,
\begin{equation}
\nonumber \lim_{n\rightarrow\infty}n^{a}P\left(\|\nu_{n,a}^{*}-\nu_{a}\|>n^{-\xi}\right)=0.
\end{equation}
\label{L8}
\end{lemma}
\vspace{-2em}
\noindent
\textit{Proof.}
We first show the present lemma (a). Since $\nu_{n,a}$ contains multiplications of possibly different components of $\Psi_{n}=\sqrt{n}(S_{n}-S)$, it suffices to show the result for $s_{i}(\beta_{0})$, the least favorable term with respect to the value of $q_{2}$ in $f(X_{i},\beta_{0})$. Let $\bar{s}_{n}(\beta)=n^{-1}\sum_{i=1}^{n}s_{i}(\beta)$ and write $\bar{s}_{n}\equiv \bar{s}_{n}(\beta_{0})$. Then the least favorable term in $\Psi_{n}$ is $\sqrt{n}(\bar{s}_{n}-Es_{i})$.  Thus,
\begin{equation}
n^{\alpha(m)}E\Pi_{\mu=1}^{m}\Psi_{n,j_{\mu}} = n^{\alpha(m)-\frac{m}{2}}E\left(\sum_{i=1}^{n}\left(s_{i}-Es_{i}\right)\right)^{m}
\end{equation}
for $2\leq m \leq 2a+2$. By expanding the RHS for each $m$ and by Assumption \ref{A1}, we can find the least favorable moment in $\nu_{n,a}$. In addition, by taking the limit, we can find $\nu_{a}$. In order for all the quantities to be well defined, the most restrictive case is the existence of $Es_{i}^{2a+2}$. For EL or ET, $s_{i}=\alpha_{0}\cdot\rho_{j}^{2}(\lambda_{0}'g_{i0})\cdot g_{0}^{k_{0}}\Pi_{l=1}^{d+1} G_{0}^{(l)k_{l}}$, $1\leq j\leq d+1$, where $\alpha_{0}$ denotes products of components of $\beta_{0}$. Since $\rho_{j}(\nu)=(\partial^{j})(\partial\nu^{j})\log(1-\nu)$, $1\leq j\leq d+1$ for EL, $Es_{i}^{2a+2}$ exists and finite under Assumptions 2-3. In particular, UBC \eqref{UBC} ensures that $E|\rho_{j}(\lambda_{0}'g_{i0})|^{k_{\rho}}<\infty$ for any finite $k_{\rho}$ and for $j=1,...,d+1$. For ET, $\rho_{j}(\nu)=-e^{\nu}$ for $1\leq j\leq d+1$. Thus, $s_{i}=\alpha_{0}\cdot e^{2\lambda_{0}g_{i0}}\cdot g_{0}^{k_{0}}\Pi_{l=1}^{d+1} G_{0}^{(l)k_{l}}$, for $1\leq j\leq d+1$. This case is not trivial. By H\"{o}lder's inequality, a sufficient condition for $Es_{i}^{2a+2}$ to exist and finite is $q_{2}> 4(a+1)$. Note that the values of $k_{0}$ and $k_{l}$'s do not matter as long as they are finite. A similar argument applies to ETEL.

Next we show the present lemma (b). Since the bootstrap sample is iid, the proof is analogous to that of the present lemma (a) by replacing $E$, $X_{i}$, and $\beta_{0}$ with $E^{*}$, $X_{i}^{*}$, and $\hat{\beta}$, respectively. We describe the proof with $m=2$, and this illustrates the proof for other values of $m$. Since $n^{\alpha(2)}=1$, $\nu_{n,a}^{*}$ contains moments of the form
\begin{equation}
\nonumber n^{\alpha(2)}E^{*}\Pi_{\mu=1}^{2}\Psi_{n,j_{\mu}}^{*}=E^{*}\hat{s}_{i}^{*2}-(E^{*}\hat{s}_{i}^{*})^{2}=n^{-1}\sum_{i=1}^{n}\hat{s}_{i}^{2}-\left(n^{-1}\sum_{i=1}^{n}\hat{s}_{i}\right)^{2}.
\end{equation}
Since the corresponding moment in $\nu_{a}$ is $Es_{i}^{2}-(Es_{i})^{2}$, combining the following results proves the lemma for $m=2$:
\begin{equation}
P\left(\left\|n^{-1}\sum_{i=1}^{n}(\hat{u}_{i}-u_{i})\right\|>n^{-\xi}\right)=P\left(\left\|n^{-1}\sum_{i=1}^{n}u_{i}-Eu_{i}\right\|>n^{-\xi}\right)=o(n^{-a}),
\label{03261}
\end{equation}
where $\hat{u}_{i}=\hat{s}_{i}$ or $\hat{u}_{i}=\hat{s}_{i}^{2}$, and $u_{i}=s_{i}$ or $u_{i}=s_{i}^{2}$. We use the fact $\|\hat{s}_{i}^{2}-s_{i}^{2}\|\leq\|\hat{s}_{i}-s_{i}\|(\|\hat{s}_{i}-s_{i}\|+2s_{i})$, the proof of \eqref{lem101}, AL1(b), and Lemma \ref{L3} to show the first result of \eqref{03261}. The second result is shown by AL1(a) with $c=\xi$ and $h(X_{i})=s_{i}^{2}-Es_{i}^{2}$ or $h(X_{i})=s_{i}-Es_{i}$. For other values of $m$, we can show \eqref{03261} for $u_{i}=s_{i}^{m}$ by using the binomial expansion, AL1, Lemma \ref{L3}, and the proof of \eqref{lem101}. The most restrictive condition arises when we apply AL1(a) with $c=\xi$ and $h(X_{i})=s_{i}^{2a+2}-Es_{i}^{2a+2}$, and we need $q_{2}>\max\left\{8(a+1),8a(a+1)(1-2\xi)^{-1}\right\}$ by H\"{o}lder's inequality. \qed

\vspace{1em}

\begin{lemma}
(a) Suppose Assumptions 1-4 hold with $q_{1}\geq 2$, $q_{1}>2a$, and $q_{2}>\max\left\{4(a+1),\frac{2ad}{d-2a-1}\right\}$ and $d\geq 2a+2$ for some $a\geq 0$, where $2a$ is a positive integer. Then,
\begin{eqnarray}
\nonumber &&\lim_{n\rightarrow\infty}n^{a}\sup_{z\in\mathbf{R}}\left|P(T_{MR}\leq z)-\left[1+\sum_{i=1}^{2a}n^{-i/2}\pi_{i}(\delta,\nu_{a})\right]\Phi(z)\right|=0,\text{ and}\\
\nonumber &&\lim_{n\rightarrow\infty}n^{a}\sup_{z\in\mathbf{R}}\left|P(\mathcal{W}_{MR}\leq z)-\int_{-\infty}^{z}d\left[1+\sum_{i=1}^{[a]}n^{-i}\pi_{\mathcal{W},i}(y,\nu_{a})\right]P(\chi^{2}_{L_{\eta}}\leq y)\right|=0.
\end{eqnarray}
(b) Suppose Assumptions 1-4 hold with $q_{1}\geq 2$, $q_{1}>4a$, and $q_{2}>\max\left\{8(a+1),8a(a+1),\frac{4ad}{d-2a-1}\right\}$ and $d\geq 2a+2$ for some $a\geq 0$, where $2a$ is a positive integer. Then,
\begin{eqnarray}
\nonumber &&\lim_{n\rightarrow\infty}n^{a}P\left(\sup_{z\in\mathbf{R}}\left|P^{*}(T_{MR}^{*}\leq z)-\left[1+\sum_{i=1}^{2a}n^{-i/2}\pi_{i}(\delta,\nu_{n,a}^{*})\right]\Phi(z)\right|>n^{-a}\right)=0,\text{ and}\\
\nonumber && \lim_{n\rightarrow\infty}n^{a}P\left(\sup_{z\in\mathbf{R}}\left|P^{*}(\mathcal{W}_{MR}^{*}\leq z)-\int_{-\infty}^{z}d\left[1+\sum_{i=1}^{[a]}n^{-i}\pi_{\mathcal{W},i}(y,\nu_{n,a}^{*})\right]P(\chi^{2}_{L_{\eta}}\leq y)\right|>n^{-a}\right)=0.
\end{eqnarray}
\label{L9}
\end{lemma}
\vspace{-1em}
\noindent
\textit{Proof.} The proof is analogous to that of Lemma 16 of A2002. We use our Lemma \ref{L7} instead of his Lemma 13. The coefficients $\nu_{a}$ are well defined by Lemma \ref{L8}(a). Lemma \ref{L8}(b) with $\xi=0$ ensures that the coefficients $\nu^{*}_{n,a}$ are well behaved. \qed

\subsection{\normalsize{Proof of Theorem 1}}

\textit{Proof.} We use Lemmas \ref{L8}-\ref{L9} to show the present Theorem. For part (a), let $a=1$. Then $q_{1}>4$, $q_{2}>\max\{16(1-2\xi)^{-1},4d(d-3)^{-1}\}$, and $d\geq4$. Since $16(1-2\xi)^{-1}\geq4d(d-3)^{-1}$ and $4d(d-3)^{-1}$ is decreasing for all $d\geq4$ and all $\xi>0$, $q_{2}>16(1-2\xi)^{-1}$ is sufficient. We first show \eqref{outline1}. By the triangle inequality,
\begin{eqnarray}
&& P\left(\sup_{z\in\mathbf{R}}\left|P(T_{MR}\leq z)-P^{*}(T_{MR}^{*}\leq z)\right|>n^{-(1/2+\xi)}\varepsilon\right)
\label{onesideT}\\
\nonumber &\leq& P\left(\sup_{z\in\mathbf{R}}\left|P(T_{MR}\leq z)-\left(1+\sum_{i=1}^{2}n^{-i/2}\pi_{i}(\delta,\nu_{1})\right)\Phi(z)\right|>n^{-(1/2+\xi)}\frac{\varepsilon}{4}\right)\\
\nonumber && + P\left(\sup_{z\in\mathbf{R}}\left|P^{*}(T_{MR}^{*}\leq z)-\left(1+\sum_{i=1}^{2}n^{-i/2}\pi_{i}(\delta,\nu_{n,1}^{*})\right)\Phi(z)\right|>n^{-(1/2+\xi)}\frac{\varepsilon}{4}\right)\\
\nonumber && + P\left(\sup_{z\in\mathbf{R}}n^{-1/2}\left|\pi_{1}(\delta,\nu_{1})-\pi_{1}(\delta,\nu_{n,1}^{*})\right|\Phi(z)>n^{-(1/2+\xi)}\frac{\varepsilon}{4}\right)\\
\nonumber && + P\left(\sup_{z\in\mathbf{R}}n^{-1}\left|\pi_{2}(\delta,\nu_{1})-\pi_{2}(\delta,\nu_{n,1}^{*})\right|\Phi(z)>n^{-(1/2+\xi)}\frac{\varepsilon}{4}\right) = o(n^{-1}).
\end{eqnarray}
The last equality holds by Lemma \ref{L9}(a)-(b) and Lemma \ref{L8}(b). The rest of the proof follows the same argument with (5.32)-(5.34) in the proof of Theorem 2 of Andrews (2001). This establishes the first result of the present theorem (a). The second result can be proved analogously.

For part (b), let $a=3/2$. Then, we need $q_{1}>6$, $q_{2}>\max\{30(1-\xi)^{-1},6d(d-4)^{-1}\}$, and $d\geq5$. Since $30(1-2\xi)^{-1}\geq6d(d-4)^{-1}$ and $6d(d-4)^{-1}$ is decreasing for all $d\geq5$ and all $\xi>0$, $q_{2}>30(1-2\xi)^{-1}$ is sufficient. For the $t$ test, we use the evenness of $\pi_{i}(\delta,\nu_{3/2})\Phi(z)$ and $\pi_{i}(\delta,\nu_{n,3/2}^{*})\Phi(z)$ for $i=1,3$ to cancel out these terms through $\Phi(z)-\Phi(-z)$. For the Wald test, there is only one expansion term because $[a]=1$. The rest follows analogously. 

For part (c), let $a=2$. To use Lemma \ref{L9}, we need $q_{1}>8$, $q_{2}>\max\{48,8d(d-5)^{-1}\}$, and $d\geq6$. The proof is the same with that of Theorem 2(c) of A2002 with his Lemmas 13 and 16 replaced by our Lemmas \ref{L7} and \ref{L9}. It relies on the argument of Hall (1988, 1992)'s methods developed for ``smooth functions of sample averages,'' for iid data. Hall's proof is quite involved, and it implicitly assumes that $\|\nu_{n,2}^{*}-\nu_{2}\|=O_{p}(n^{-1/2})$ on a set with probability $1-o(n^{-2})$, which is stronger than our Lemma \ref{L8}(b). To use his result, I assume $q_{2}$ is sufficiently large, as is also assumed in Hall (1992, Theorem 5.1).\footnote{For GMM, A2002 and Lee (2014) assume sufficiently large $q_{g}$ and $q_{G}$. Note that the assumptions on $q_{2}$, $q_{g}$, and $q_{G}$, are sufficient, rather than necessary. Thus, it is possible that the result holds under smaller values of $q_{2}$, $q_{g}$, and $q_{G}$.} \qed

\section*{References}

\begin{enumerate}[leftmargin=0 in]
	\item[] Allen, J., Gregory, A. W., and Shimotsu, K. (2011). Empirical likelihood block bootstrapping. Journal of Econometrics, 161(2), 110-121.
	\item[] Almeida, C., and Garcia, R. (2012). Assessing misspecified asset pricing models with empirical likelihood estimators. Journal of Econometrics, 170(2), 519-537.
	\item[] Altonji, J. G., and Segal, L. M. (1996). Small-sample bias in GMM estimation of covariance structures. Journal of Business and Economic Statistics, 14(3), 353-366.
	\item[] Anatolyev, S. (2005). GMM, GEL, serial correlation, and asymptotic bias. Econometrica, 73(3), 983-1002.
	\item[] Andrews, D. W.  (2001). Higher-order improvements of a computationally attractive k-step bootstrap for extremum estimators. Cowles Foundation Discussion Paper No. 1269, Yale University. Available at \href{http://cowles.econ.yale.edu}{http://cowles.econ.yale.edu.}
	\item[] Andrews, D. W. (2002). Higher-order improvements of a computationally attractive k-step bootstrap for extremum estimators. Econometrica, 70(1), 119-162.
	\item[] Antoine, B., Bonnal, H., and Renault, E. (2007). On the efficient use of the informational content of estimating equations: Implied probabilities and Euclidean empirical likelihood. Journal of Econometrics, 138(2), 461-487.
	\item[] Beran, R. (1988). Prepivoting test statistics: a bootstrap view of asymptotic refinements. Journal of the American Statistical Association, 83(403), 687-697.
	\item[] Bickel, P. J., and Freedman, D. A. (1981). Some asymptotic theory for the bootstrap. The Annals of Statistics, 9(6), 1196-1217.
	\item[] Blundell, R., and Bond, S. (1998). Initial conditions and moment restrictions in dynamic panel data models. Journal of Econometrics, 87(1), 115-143.
	\item[] Bond, S., and Windmeijer, F. (2005). Reliable inference for GMM estimators? Finite sample properties of alternative test procedures in linear panel data models. Econometric Reviews, 24(1), 1-37.
	\item[] Bravo, F. (2010). Efficient M-estimators with auxiliary information. Journal of Statistical Planning and Inference, 140(11), 3326-3342.
	\item[] Brown, B. W., and Newey, W. K. (2002). Generalized method of moments, efficient bootstrapping, and improved inference. Journal of Business and Economic Statistics, 20(4), 507-517.
	\item[] Canay, I. A. (2010). EL inference for partially identified models: large deviations optimality and bootstrap validity. Journal of Econometrics, 156(2), 408-425.
	\item[] Chamberlain, G., and Imbens, G. W. (2003). Nonparametric applications of Bayesian inference. Journal of Business and Economic Statistics, 21(1), 12-18.
	\item[] Chen, X., Hong, H., and Shum, M. (2007). Nonparametric likelihood ratio model selection tests between parametric likelihood and moment condition models. Journal of Econometrics, 141(1), 109-140.
	\item[] Davidson, R., and MacKinnon, J. G. (2002). Fast double bootstrap tests of nonnested linear regression models. Econometric Reviews, 21(4), 419-429.
	\item[] Davidson, R., and MacKinnon, J. G. (2007). Improving the reliability of bootstrap tests with the fast double bootstrap. Computational Statistics \& Data Analysis, 51(7), 3259-3281.
	\item[] Giacomini, R., Politis, D. N., and White, H. (2013). A warp-speed method for conducting Monte Carlo experiments involving bootstrap estimators. Econometric Theory, 1-23.
	\item[] Guggenberger, P. (2008). Finite sample evidence suggesting a heavy tail problem of the generalized empirical likelihood estimator. Econometric Reviews, 27(4-6), 526-541.
	\item[] Guggenberger, P., and Hahn, J. (2005). Finite Sample Properties of the Two-Step Empirical Likelihood Estimator. Econometric Reviews, 24(3), 247-263.
	\item[] Hahn, J. (1996). A note on bootstrapping generalized method of moments estimators. Econometric Theory, 12, 187-197.
	\item[] Hall, A. R., and Inoue, A. (2003). The large sample behaviour of the generalized method of moments estimator in misspecified models. Journal of Econometrics, 114(2), 361-394.
	\item[] Hall, P. (1988). On symmetric bootstrap confidence intervals. Journal of the Royal Statistical Society. Series B (Methodological), 35-45.
	\item[] Hall, P. (1992). The bootstrap and Edgeworth expansion. Springer.
	\item[] Hall, P., and Horowitz, J. L. (1996). Bootstrap critical values for tests based on generalized-method-of-moments estimators. Econometrica, 64(4), 891-916.
	\item[] Hansen, L. P. (1982). Large sample properties of generalized method of moments estimators. Econometrica, 50(4), 1029-1054.
	\item[] Hansen, L. P., Heaton, J., and Yaron, A. (1996). Finite-sample properties of some alternative GMM estimators. Journal of Business and Economic Statistics, 14(3), 262-280.
	\item[] Hansen, L. P., and Jagannathan, R. (1997). Assessing specification errors in stochastic discount factor models. The Journal of Finance, 52(2), 557-590.
	\item[] Hellerstein, J. K., and Imbens, G. W. (1999). Imposing moment restrictions from auxiliary data by weighting. Review of Economics and Statistics, 81(1), 1-14.
	\item[] Horowitz, J. L. (2001). The bootstrap. Handbook of Econometrics, 5, 3159-3228.
	\item[] Imbens, G. W. (1997). One-step estimators for over-identified generalized method of moments models. The Review of Economic Studies, 64(3), 359-383.
	\item[] Imbens, G. W. (2002). Generalized method of moments and empirical likelihood. Journal of Business and Economic Statistics, 20(4).
	\item[] Imbens, G. W., Spady, R. H. and Johnson, P. (1998). Information theoretic approaches to inference in moment condition models. Econometrica, 66(2), 333-357.
	\item[] Inoue, A., and Shintani, M. (2006). Bootstrapping GMM estimators for time series. Journal of Econometrics, 133(2), 531-555.
	\item[] Kitamura, Y., and Stutzer, M. (1997). An information-theoretic alternative to generalized method of moments estimation. Econometrica, 65(4), 861-874.
	\item[] Kitamura, Y., Otsu, T., and Evdokimov, K. (2013). Robustness, infinitesimal neighborhoods, and moment restrictions. Econometrica, 81(3), 1185-1201.
	\item[] Kundhi, G., and Rilstone, P. (2012). Edgeworth expansions for GEL estimators. Journal of Multivariate Analysis, 106, 118-146.
	\item[] Lee, S. (2014). Asymptotic refinements of a misspecification-robust bootstrap for generalized method of moments estimators. Journal of Econometrics, 178(3), 398-413.
	\item[] Lee, S. (2014b). On robustness of GEL estimators to model misspecification. Working paper. UNSW Business School, University of New South Wales.
	\item[] Newey, W. K., and McFadden, D. (1994). Large sample estimation and hypothesis testing. Handbook of Econometrics, 4, 2111-2245.
	\item[] Newey, W. K., and Smith, R. J. (2004). Higher order properties of GMM and generalized empirical likelihood estimators. Econometrica, 72(1), 219-255.
	\item[] Owen, A. B. (1988). Empirical likelihood ratio confidence intervals for a single functional. Biometrika, 75(2), 237-249.
	\item[] Owen, A. (1990). Empirical likelihood ratio confidence regions. The Annals of Statistics, 18(1), 90-120.
	\item[] Qin, J., and Lawless, J. (1994). Empirical likelihood and general estimating equations. The Annals of Statistics, 300-325.
	\item[] Schennach, S. M. (2007). Point estimation with exponentially tilted empirical likelihood. The Annals of Statistics, 35(2), 634-672.
	\item[] White, H. (2000). A reality check for data snooping. Econometrica, 68(5), 1097-1126.
	\item[] Windmeijer, F. (2005). A finite sample correction for the variance of linear efficient two-step GMM estimators. Journal of Econometrics, 126(1), 25-51.
\end{enumerate}

\newpage
\thispagestyle{empty}

\begin{table}[p!]
\centering
\begin{tabular}{cllllrrrrrr}
\toprule
\multirow{3}{*}{DGP C-1} & & & & & \multicolumn{3}{c}{$n=100$}  & \multicolumn{3}{c}{$n=200$} \\
\cmidrule(r){6-11}
& & & & & \multicolumn{2}{c}{CI}  & J test & \multicolumn{2}{c}{CI}  & J test\\
& & & & & .90 & .95 					  & .05     & .90 & .95 				   & .05\\
\toprule
\multirow{16}{*}{T=4} & \multirow{8}{*}{Boot} & GMM & C & HH  & .926 & .973 & .006 & .922 & .967 & .024\\
            					&  								 & GMM & MR & L & .941 & .981 				 & n/a   & .941 & .976 & n/a \\
            					&  								 & EL 	  & MR& L & .929 & .975 & \multirow{2}{*}{n/a}& .923 & .973 & \multirow{2}{*}{n/a}\\
            					&  								 & EL 	  & MR & BN  & .885 & .940 & 								& .885 & .943 & \\
            					&  								 & ET 	  & MR & L & .926 & .976 & \multirow{2}{*}{n/a} & .921 & .974 & \multirow{2}{*}{n/a}\\
            					&  								 & ET 	  & MR & BN  & .897 & .950 & 								 & .897 & .953 & \\
						        &     						  	& ETEL  & MR & L & .928 & .976 & \multirow{2}{*}{n/a} & .920 & .972 & \multirow{2}{*}{n/a}\\
						        &     						  	& ETEL  & MR & BN  & .898 & .947 & 							   & .892 & .951 & \\
\cmidrule(r){2-11}
                 		 & \multirow{8}{*}{Asymp} & GMM & MR & & .779 & .846 & \multirow{2}{*}{.038} & .829 & .895 & \multirow{2}{*}{.042}\\
                 		 & 									  & GMM & C  & & .770 & .844 & 								 				& .827 & .893 &\\
                 		 & 									  & EL 	   & MR& & .734 & .807 & \multirow{2}{*}{.125} & .814 & .877 & \multirow{2}{*}{.088}\\
                 		 & 									  & EL 	   & C   & & .724 & .802  &  							 		& .797 & .867 &\\
          		         & 									  & ET 	   & MR & & .747 & .819 & \multirow{2}{*}{.103} & .824 & .884 & \multirow{2}{*}{.079}\\
                 		 & 									  & ET     & C    & & .724 & .810 &  							  		 & .803 & .869 &\\
                 		 & 									  & ETEL  & MR &  & .742 & .816 & \multirow{2}{*}{.172} & .814 & .878 & \multirow{2}{*}{.117}\\
                 		 & 									  & ETEL  & C   &  & .736 & .815 &  								 		 & .805 & .872 &\\
\midrule
\multirow{16}{*}{T=6} & \multirow{8}{*}{Boot} & GMM & C & HH  & .950 & .983 & .000& .932 & .975 & .002\\
            					&  								 	  & GMM & MR & L & .971 & .990 & n/a & .950 & .987 & n/a\\
            					&  								 	  & EL 	  & MR   & L & .967 & .991 & \multirow{2}{*}{n/a}& .934 & .974 & \multirow{2}{*}{n/a}\\
            					&  								 	  & EL 	  & MR & BN  & .922 & .966 & & .911 & .957 & \\
            					&  								 	  & ET 	  & MR  & L & .959 & .987 & \multirow{2}{*}{n/a}& .925 & .973 & \multirow{2}{*}{n/a}\\
            					&  								 	  & ET 	  & MR & BN  & .928 & .970 & & .904 & .953 & \\
						        &     						  		  & ETEL  & MR  & L & .958 & .986 & \multirow{2}{*}{n/a}& .926 & .973 & \multirow{2}{*}{n/a}\\
						        &     						  		  & ETEL  & MR & BN  & .917 & .965 & & .912 & .954 & \\
\cmidrule(r){2-11}
                 		 & \multirow{8}{*}{Asymp} & GMM & MR & & .661 & .742 & \multirow{2}{*}{.040} & .760 & .835 & \multirow{2}{*}{.045}\\
                 		 & 									  & GMM & C  & & .648 & .728 & 						   & .759 & .836 &\\
                 		 & 									  & EL 	   & MR&  & .690 & .766 & \multirow{2}{*}{.426} & .784 & .862 & \multirow{2}{*}{.257}\\
                 		 & 									  & EL 	   & C   & & .651 & .735  &  					  & .748 & .828 &\\
          		         & 									  & ET 	   & MR & & .724 & .799 & \multirow{2}{*}{.339} & .808 & .878 & \multirow{2}{*}{.210}\\
                 		 & 									  & ET     & C    & & .655 & .737 &  						& .761 & .841 &\\
                 		 & 									  & ETEL  & MR &  & .708 & .780 & \multirow{2}{*}{.568} & .794 & .871 & \multirow{2}{*}{.356}\\
                 		 & 									  & ETEL  & C   &  & .657 & .740 &  							& .754 & .839 &\\
\bottomrule
\end{tabular}
\caption{Coverage Probabilities of 90\% and 95\% Confidence Intervals for $\rho_{0}$ based on GMM, EL, ET, and ETEL under DGP C-1.}
\label{table_C1}
\end{table}

\clearpage
\thispagestyle{empty}

\begin{table}[p!]
\centering
\begin{tabular}{cllllrrrrrr}
\toprule
\multirow{3}{*}{DGP C-2} & & & & & \multicolumn{3}{c}{$n=100$}  & \multicolumn{3}{c}{$n=200$} \\
\cmidrule(r){6-11}
& & & &  & \multicolumn{2}{c}{CI}  & J test & \multicolumn{2}{c}{CI}  & J test\\
& & & & & .90 & .95 					  & .05     & .90 & .95 				   & .05\\
\toprule
\multirow{16}{*}{T=4} & \multirow{8}{*}{Boot} & GMM & C& HH  & .911 & .957 & .031 & .902 & .953 & .035\\
            					&  								 & GMM & MR & L & .936 & .973 				 & n/a   & .915 & .961 & n/a \\
            					&  								 & EL 	  & MR & L & .924 & .968 & \multirow{2}{*}{n/a}& .904 & .958 & \multirow{2}{*}{n/a}\\
            					&  								 & EL 	  & MR & BN  & .872 & .936 & 								& .893 & .944 & \\
            					&  								 & ET 	  & MR & L & .922 & .967 & \multirow{2}{*}{n/a} & .902 & .960 & \multirow{2}{*}{n/a}\\
            					&  								 & ET 	  & MR & BN  & .892 & .952 & 								 & .888 & .942 & \\
						        &     						  	& ETEL  & MR & L & .921 & .967 & \multirow{2}{*}{n/a} & .903 & .957 & \multirow{2}{*}{n/a}\\
						        &     						  	& ETEL  & MR & BN  & .892 & .949 & 							   & .884 & .944 & \\
\cmidrule(r){2-11}
                 		 & \multirow{8}{*}{Asymp} & GMM & MR & & .815 & .877 & \multirow{2}{*}{.050} & .849 & .906 & \multirow{2}{*}{.049}\\
                 		 & 									  & GMM & C  & & .805 & .870 & 								 				& .847 & .904 &\\
                 		 & 									  & EL 	   & MR & & .793 & .857 & \multirow{2}{*}{.090} & .842 & .901 & \multirow{2}{*}{.066}\\
                 		 & 									  & EL 	   & C  &  & .786 & .850  &  							 		& .834 & .895 &\\
          		         & 									  & ET 	   & MR & & .797 & .864 & \multirow{2}{*}{.087} & .844 & .904 & \multirow{2}{*}{.067}\\
                 		 & 									  & ET     & C    & & .783 & .847 &  							  		 & .836 & .893 &\\
                 		 & 									  & ETEL  & MR &  & .797 & .858 & \multirow{2}{*}{.114} & .844 & .902 & \multirow{2}{*}{.079}\\
                 		 & 									  & ETEL  & C   &  & .789 & .853 &  								 		 & .836 & .896 &\\
\midrule
\multirow{16}{*}{T=6} & \multirow{8}{*}{Boot} & GMM & C & HH  & .936 & .973 & .006 & .905 & .953 & .023\\
            					&  								 	  & GMM & MR & L & .970 & .990 & n/a & .940 & .974 & n/a\\
            					&  								 	  & EL 	  & MR  & L & .965 & .988 & \multirow{2}{*}{n/a}& .917 & .970 & \multirow{2}{*}{n/a}\\
            					&  								 	  & EL 	  & MR & BN  & .911 & .963 & & .896 & .950 & \\
            					&  								 	  & ET 	  & MR   & L & .952 & .985 & \multirow{2}{*}{n/a}& .910 & .963 & \multirow{2}{*}{n/a}\\
            					&  								 	  & ET 	  & MR & BN  & .926 & .962 & & .896 & .948 & \\
						        &     						  		  & ETEL  & MR  & L & .956 & .986 & \multirow{2}{*}{n/a}& .914 & .966 & \multirow{2}{*}{n/a}\\
						        &     						  		  & ETEL  & MR & BN  & .915 & .959 & & .895 & .948 & \\
\cmidrule(r){2-11}
                 		 & \multirow{8}{*}{Asymp} & GMM & MR & & .716 & .792 & \multirow{2}{*}{.049} &.801  & .868 & \multirow{2}{*}{.050}\\
                 		 & 									  & GMM & C  & & .715 & .797 & 						   & .805 & .874 &\\
                 		 & 									  & EL 	   & MR & & .756 & .826 & \multirow{2}{*}{.281} & .837 & .900 & \multirow{2}{*}{.152}\\
                 		 & 									  & EL 	   & C   & & .730 & .813  &  					  & .818 & .883 &\\
          		         & 									  & ET 	   & MR & & .782 & .846 & \multirow{2}{*}{.251} & .843 & .909 & \multirow{2}{*}{.152}\\
                 		 & 									  & ET     & C    & & .741 & .812 &  						& .820 & .888 &\\
                 		 & 									  & ETEL  & MR &  & .768 & .838 & \multirow{2}{*}{.383} & .842 & .903 & \multirow{2}{*}{.210}\\
                 		 & 									  & ETEL  & C    & & .742 & .820 &  							& .825 & .886 &\\
\bottomrule
\end{tabular}
\caption{Coverage Probabilities of 90\% and 95\% Confidence Intervals for $\rho_{0}$ based on GMM, EL, ET, and ETEL under DGP C-2.}
\label{table_C2}
\end{table}

\clearpage
\thispagestyle{empty}

\begin{table}[p!]
\centering
\begin{tabular}{cllllrrrrrr}
\toprule
\multirow{3}{*}{DGP M-1} & & & & & \multicolumn{3}{c}{$n=100$}  & \multicolumn{3}{c}{$n=200$} \\
\cmidrule(r){6-11}
& & & & &  \multicolumn{2}{c}{CI}  & J test & \multicolumn{2}{c}{CI}  & J test\\
& & & & & .90 & .95 					  & .05     & .90 & .95 				   & .05\\
\toprule
\multirow{16}{*}{T=4} & \multirow{8}{*}{Boot} & GMM & C & HH  & .841 & .938 & .003 & .882 & .945 & .041\\
            					&  								 & GMM & MR & L & .919 & .967 				 & n/a   & .949 & .982 & n/a \\
            					&  								 & EL 	  & MR & L & .826 & .891 & \multirow{2}{*}{n/a}& .854 & .923 & \multirow{2}{*}{n/a}\\
            					&  								 & EL 	  & MR & BN  & .755 & .830 & 								& .797 & .863 & \\
            					&  								 & ET 	  & MR & L & .833 & .896 & \multirow{2}{*}{n/a} & .868 & .930 & \multirow{2}{*}{n/a}\\
            					&  								 & ET 	  & MR & BN  & .761 & .842 & 								 & .797 & .870 & \\
						        &     						  	& ETEL  & MR & L & .824 & .887 & \multirow{2}{*}{n/a} & .851 & .922 & \multirow{2}{*}{n/a}\\
						        &     						  	& ETEL  & MR & BN  & .759 & .830 & 							   & .796 & .861 & \\
\cmidrule(r){2-11}
                 		 & \multirow{8}{*}{Asymp} & GMM & MR &  & .522 & .575 & \multirow{2}{*}{.167} & .629 & .689 & \multirow{2}{*}{.292}\\
                 		 & 									  & GMM & C  & & .436 & .490 & 								 				& .540 & .607 &\\
                 		 & 									  & EL 	   & MR & & .595 & .659 & \multirow{2}{*}{.249} & .690 & .753 & \multirow{2}{*}{.310}\\
                 		 & 									  & EL 	   & C   & & .570 & .626  &  							 		& .629 & .702 &\\
          		         & 									  & ET 	   & MR & & .601 & .663 & \multirow{2}{*}{.228} & .698 & .766 & \multirow{2}{*}{.315}\\
                 		 & 									  & ET     & C    & & .560 & .624 &  							  		 & .622 & .699 &\\
                 		 & 									  & ETEL  & MR &  & .608 & .670 & \multirow{2}{*}{.307} & .703 & .766 & \multirow{2}{*}{.368}\\
                 		 & 									  & ETEL  & C    & & .587 & .643 &  								 		 & .642 & .715 &\\
\midrule
\multirow{16}{*}{T=6} & \multirow{8}{*}{Boot} & GMM & C & HH  & .916 & .969 & .000 & .940 & .977 & .009\\
            					&  								 	  & GMM & MR & L & .972 & .992 & n/a & .987 & .996 & n/a\\
            					&  								 	  & EL 	  & MR  & L & .936 & .973 & \multirow{2}{*}{n/a}& .933 & .972 & \multirow{2}{*}{n/a}\\
            					&  								 	  & EL 	  & MR & BN  & .826 & .889 & & .831 & .890 & \\
            					&  								 	  & ET 	  & MR  & L & .934 & .969 & \multirow{2}{*}{n/a}& .935 & .976 & \multirow{2}{*}{n/a}\\
            					&  								 	  & ET 	  & MR & BN  & .853 & .912 & & .843 & .914 & \\
						        &     						  		  & ETEL  & MR & L & .929 & .971 & \multirow{2}{*}{n/a}& .926 & .969 & \multirow{2}{*}{n/a}\\
						        &     						  		  & ETEL  & MR & BN  & .829 & .895 & & .821 & .886 & \\
\cmidrule(r){2-11}
                 		 & \multirow{8}{*}{Asymp} & GMM & MR & & .429 & .488 & \multirow{2}{*}{.253} & .575 & .651 & \multirow{2}{*}{.604}\\
                 		 & 									  & GMM & C  & & .335 & .393 & 						   & .491 & .562 &\\
                 		 & 									  & EL 	   & MR  & & .579 & .651 & \multirow{2}{*}{.801} & .693 & .766 & \multirow{2}{*}{.879}\\
                 		 & 									  & EL 	   & C   & & .484 & .551  &  					  & .542 & .624 &\\
          		         & 									  & ET 	   & MR & & .629 & .695 & \multirow{2}{*}{.739} & .749 & .810 & \multirow{2}{*}{.860}\\
                 		 & 									  & ET     & C    & & .483 & .554 &  						& .564 & .647 &\\
                 		 & 									  & ETEL  & MR &  & .599 & .665 & \multirow{2}{*}{.884} & .713 & .783 & \multirow{2}{*}{.932}\\
                 		 & 									  & ETEL  & C    & & .485 & .559 &  							& .548 & .631 &\\
\bottomrule
\end{tabular}
\caption{Coverage Probabilities of 90\% and 95\% Confidence Intervals for $\rho_{0}$ based on GMM, EL, ET, and ETEL under DGP M-1.}
\label{table_M1}
\end{table}

\clearpage
\thispagestyle{empty}

\begin{table}[p!]
\centering
\begin{tabular}{cllllcccccc}
\toprule
\multirow{3}{*}{DGP M-2} & & & & & \multicolumn{3}{c}{$n=100$}  & \multicolumn{3}{c}{$n=200$} \\
\cmidrule(r){6-11}
& & & & &  \multicolumn{2}{c}{CI}  & J test & \multicolumn{2}{c}{CI}  & J test \\
& & & & & .90 & .95 					  & .05     & .90 & .95 				   & .05\\
\toprule
\multirow{16}{*}{T=4} & \multirow{8}{*}{Boot} & GMM & C & HH  & .857 & .933 & .000 & .861 & .942 & .008\\
            					&  								 				& GMM & MR & L & .921 & .969 				 & n/a   & .933 & .977 & n/a \\
            					&  								 				& EL 	  & MR & L & .835 & .926 & \multirow{2}{*}{n/a}& .826 & .917 & \multirow{2}{*}{n/a}\\
            					&  								 				& EL 	  & MR & BN  & .747 & .826 & 								& .732 & .820 & \\
            					&  								 				& ET 	  & MR & L & .839 & .930 & \multirow{2}{*}{n/a} & .844 & .925 & \multirow{2}{*}{n/a}\\
            					&  								 				& ET 	  & MR & BN  & .735 & .823 & 								 & .735 & .821 & \\
						        &     						  					& ETEL  & MR & L & .823 & .914 & \multirow{2}{*}{n/a} & .824 & .908 & \multirow{2}{*}{n/a}\\
						        &     						  					& ETEL  & MR & BN  & .737 & .820 & 							   & .733 & .819 & \\
\cmidrule(r){2-11}
                 		 & \multirow{8}{*}{Asymp} & GMM & MR & & .562 & .626 & \multirow{2}{*}{.129} & .629 & .695 & \multirow{2}{*}{.262}\\
                 		 & 									  & GMM & C  & & .510 & .573 & 								 				& .580 & .652 &\\
                 		 & 									  & EL 	   & MR & & .533 & .597 & \multirow{2}{*}{.316} & .578 & .642 & \multirow{2}{*}{.417}\\
                 		 & 									  & EL 	   & C   & & .500 & .569  &  							 		& .520 & .589 &\\
          		         & 									  & ET 	   & MR & & .553 & .618 & \multirow{2}{*}{.262} & .592 & .660 & \multirow{2}{*}{.393}\\
                 		 & 									  & ET     & C    & & .506 & .574 &  							  		 & .527 & .593 &\\
                 		 & 									  & ETEL  & MR &  & .545 & .608 & \multirow{2}{*}{.399} & .592 & .661 & \multirow{2}{*}{.491}\\
                 		 & 									  & ETEL  & C   &  & .530 & .601 &  								 		 & .539 & .607 &\\
\midrule
\multirow{16}{*}{T=6} & \multirow{8}{*}{Boot} & GMM & C & HH  	& .941 & .975 & .000 & .903 & .973 & .000\\
            					&  								 	  			& GMM & MR & L	 & .978 & .991 & n/a & .989 & .994 & n/a\\
            					&  								 	  			& EL 	  & MR   & L & .951 & .982 & \multirow{2}{*}{n/a}& .926 & .972 & \multirow{2}{*}{n/a}\\
            					&  								 	  			& EL 	  & MR & BN    & .826 & .895 & & .800 & .878 & \\
            					&  								 	  			& ET 	  & MR   & 	L		& .943 & .979 & \multirow{2}{*}{n/a}& .926 & .975 & \multirow{2}{*}{n/a}\\
            					&  								 	  			& ET 	  & MR & BN    & .831 & .899 & & .834 & .899 & \\
						        &     						  		  			& ETEL  & MR  &  	L	  & .928 & .973 & \multirow{2}{*}{n/a}& .922 & .970 & \multirow{2}{*}{n/a}\\
						        &     						  		  			& ETEL  & MR & BN    & .816 & .876 & & .798 & .869 & \\
\cmidrule(r){2-11}
		                 		 & \multirow{8}{*}{Asymp} & GMM & MR &  & .457 & .517 & \multirow{2}{*}{.241} & .530 & .603 & \multirow{2}{*}{.633}\\
        		         		 & 									  		& GMM & C  & 	 & .388 & .452 & 						   & .443 & .515 &\\
                		 		 & 									  		& EL 	   & MR  & & .534 & .608 & \multirow{2}{*}{.878} & .635 & .710 & \multirow{2}{*}{.949}\\
                 				 & 									  		& EL 	   & C   &   & .447 & .520  &  					  & .499 & .578 &\\
          		         		& 									  		& ET 	   & MR &  & .582 & .655 & \multirow{2}{*}{.801} & .688 & .764 & \multirow{2}{*}{.922}\\
                 		 		& 									  		& ET     & C    &    & .425 & .502 &  						& .506 & .582 &\\
                 		 		& 									  		& ETEL  & MR &   & .570 & .641 & \multirow{2}{*}{.943} & .664 & .733 & \multirow{2}{*}{.976}\\
                 		 		& 									  		& ETEL  & C    &   & .445 & .520 &  							& .507 & .579 &\\
\bottomrule
\end{tabular}
\caption{Coverage Probabilities of 90\% and 95\% Confidence Intervals for $\rho_{0}$ based on GMM, EL, ET, and ETEL under DGP M-2.}
\label{table_M2}
\end{table}

\clearpage
\thispagestyle{empty}

\begin{table}[p!]
\centering
\begin{tabular}{clllrrrrrrrr}
\toprule
& & & & \multicolumn{4}{c}{$T=4$} & \multicolumn{4}{c}{$T=6$}\\
DGP & & & & \multicolumn{2}{c}{$n=100$}  & \multicolumn{2}{c}{$n=200$} & \multicolumn{2}{c}{$n=100$}  & \multicolumn{2}{c}{$n=200$} \\
\cmidrule(r){5-8}\cmidrule(r){9-12}
& & & & .90 & .95 					   & .90 & .95 				   & .90 & .95 					   & .90 & .95  \\
\toprule
\multirow{8}{*}{C-1}  		   & GMM & C 	& HH  & .500 & .633    & .328  & .403 	& .306 & .388   & .194 & .235\\
            					  			  & GMM & MR & L		& .558 & .715   & .357  & .443	 & .368 & .467  & .212 & .266  \\
            					  			  & EL 	   & MR  & L  	   & .633 & .848 	& .395  & .512	 & .433 & .593  & .212 & .269 \\
            					  			  & EL 	  & MR  & BN   & .533 & .673 	& .351	& .429 	 & .342  & .433 & .193  & .238 \\
            					  			  & ET 	  & MR  & L 		& .535 & .836 	& .393  & .502	 & .406  & .537  & .204 & .258 \\
            					  			  & ET 	  & MR  & BN   & .566 & .713 	& .362  & .446 	 & .349 & .441  & .189 & .230 \\
						             		  & ETEL  & MR &  L 	  & .614 & .823   & .390  & .504   & .400 & .530  & .204 & .266 \\
						             		  & ETEL  & MR & BN  & .538 & .679   & .358  & .444   & .332 & .417  & .193  & .236 \\
\midrule
\multirow{8}{*}{C-2}  		   & GMM & C & HH  	   & .517 & .651   & .350 & .429    & .334 & .402 & .202 & .242 \\
            					  			  & GMM & MR & L		& .580 & .738    & .365 & .450	  & .404 & .501 & .229 & .275 \\
            					  			  & EL 	   & MR  & L		& .625 & .800    & .381 & .477	  & .436 & .562 & .213 & .268 \\
            					  			  & EL 	  & MR  & BN   & .527 & .655    & .369	& .448 	  & .344 & .433 & .199  & .242 \\
            					  			  & ET 	  & MR  & L 		& .625 & .797 	 & .379  & .480	  & .408 & .524 & .206 & .257 \\
            					  			  & ET 	  & MR  & BN   & .562 & .725    & .364  & .439   & .363 & .435 & .198 & .238 \\
						             		  & ETEL  & MR &  L 	  & .614 & .789    & .378  & .473	& .421 & .538 & .209 & .264 \\
						             		  & ETEL  & MR & BN  & .560 & .699    & .356  & .444	& .352 & .429 & .196  & .239 \\
\midrule
\multirow{8}{*}{M-1}  		  & GMM & C & HH  		& 1.277 & 1.898  & .936   & 1.333   & .831   & 1.126  & .505  & .684\\
            					  			  & GMM & MR & L		  & 2.383 & 3.393  & 1.655 & 2.418	& 1.591 & 2.161  & .963  & 1.324  \\
            					  			  & EL 	   & MR  & 	L 	  & .956   & 1.382 	& .667  & .976   & .802   & 1.180  & .458  & .656 \\
            					  			  & EL 	  & MR  & BN     & .732   & .975 	& .536	& .702 	   & .472   & .599    & .313  & .385 \\
            					  			  & ET 	  & MR  & L 		  & .991   & 1.397 	& .693  & .987	  & .798   & 1.174  & .439  & .605 \\
            					  			  & ET 	  & MR  & BN     & .757   & 1.044 	& .538  & .700 	   & .520   & .695    & .318  & .401 \\
						             		  & ETEL  & MR &  L	    & .914   & 1.313  & .622  & .880	 & .742   & 1.074  & .451  & .590 \\
						             		  & ETEL  & MR & BN    & .715   & .935  & .506  & .649	  & .480   & .617 	 & .314  & .387 \\
\midrule
\multirow{8}{*}{M-2}  		  & GMM & C & HH  		& .890 & 1.276 & .616 & .905  & .578 & .784  & .355 & .504 \\
            					  			  & GMM & MR &  L 	 & 1.365 & 1.970 & .939 & 1.343	& .962 & 1.365  & .556 & .787 \\
            					  			  & EL 	   & MR  & L 		 & .976 & 1.435 & .745 & 1.109 & .878 & 1.308  & .497 & .708 \\
            					  			  & EL 	  & MR  & BN    & .732 & .946 	& .565 & .732 & .490  & .634 & .314 & .399 \\
            					  			  & ET 	  & MR  & 	L	 & .994 & 1.438 & .770 & 1.080 & .851 & 1.236  & .490 & .696 \\
            					  			  & ET 	  & MR  & BN    & .717 & .946 & .562 & .719 & .525 & .674  & .343 & .432 \\
						             		  & ETEL  & MR &  	L   & 1.164 & 1.665 & .767 & .995 & .771 & 1.105  & .504 & .715 \\
						             		  & ETEL  & MR & BN   & .882 & 1.150 & .539 & .623	& .479 & .597  & .322 & .405 \\
\bottomrule
\end{tabular}
\caption{Width of 90\% and 95\% Bootstrap Confidence Intervals for $\rho_{0}$ based on GMM, EL, ET, and ETEL.}
\label{table_W1}
\end{table}

\clearpage
\thispagestyle{empty}

\begin{table}[p!]
\centering
\begin{tabular}{llccccc}
\toprule
	 										&						& OLS  		& GMM & EL	 & ET 		& ETEL  \\
\midrule
\multirow{3}{*}{const}			   & $\hat{\beta}$ &	.294 	&	-.561	& .016  & -.059  & -.023\\
										    & s.e.$_{C}$ 	   &  (.235)   	&	(.089)	& (.097) &  (.101) & (.100)\\
											& s.e.$_{MR}$     &  			 &	(.194)	 & (.109)&  (.125) & (.121)\\
											&						&			&				&			& 			& \\
\multirow{3}{*}{educ}  			   & $\hat{\beta}$  & .054		&	.056	&   .068 & .070		& .071\\
											& s.e.$_{C}$   		&  (.010)  	&	(.006)	& (.005)&  (.006)	& (.006)\\
											& s.e.$_{MR}$ 	  &   		  	&	(.018)	& (.006)&  (.009)		& (.008)\\
											&						&			&				&			& & \\
\multirow{3}{*}{exper}			  & $\hat{\beta}$	& .068 		&	.140	&  .076& .081		& .082\\
											&  s.e.$_{C}$   	&  (.025)   &	(.006)	& (.007)&  (.007)	& (.007)\\
											& s.e.$_{MR}$ 	  &   		   	&	(.022)	& (.008)&  (.011)	& (.010)\\
											&						&			&					&			& & \\
\multirow{3}{*}{exper$^{2}$}   & $\hat{\beta}$  & -.002		&	-.004	& -.002& -.002			& -.002\\
											& s.e.$_{C}$    	&  (.001)  	&	(.0002)	& (.0002)&  (.0002)  & (.0002)\\
											& s.e.$_{MR}$ 	 &   		  	&	(.0006)	& (.0002)&  (.0003)	 & (.0002)\\
											&						&			 &					&			& & \\
\multirow{3}{*}{IQ}					& $\hat{\beta}$  & .004	 	&	.007	& .005 & .006				& .005\\
											& s.e.$_{C}$    	&  (.001)  &	(.001)  &  (.001)&  (.001)	  & (.001)\\
											& s.e.$_{MR}$ 	   &   		  	&	(.002)   & (.001)&  (.002)		   & (.002)\\
											&						&			&					&			& & \\
\multirow{3}{*}{KWW}			 &$\hat{\beta}$	  & .008	 & 	-.0003	& -.002 & -.004		& -.005\\
											& s.e.$_{C}$      &  (.003)  &	(.003) 	  & (.003) &  (.003) 	   & (.003)\\
											& s.e.$_{MR}$ 	&   		  &		(.007)	& (.003)&  (.004) 		& (.004)\\
\midrule
J test									&	$\chi^{2}_{13} $&			&	477.3	& 177.5 & 285.2		& 196.2 \\
										 &	p-value			 &			 	&	[.000]	& [.000] & [.000]	   & [.000] \\	
\bottomrule
\end{tabular}
\caption{Estimation of the Mincer equation using Census moments}
\label{tb_est}
\end{table}

\clearpage
\thispagestyle{empty}

\begin{table}[p!]
\centering
\begin{tabular}{llllcccc}
\toprule
Estimator	& CI	& s.e. & & LB & Point Est. & UB	& Width  \\
\midrule
OLS			& Asymp 	& n/a & & .033 & .054 & .074 & .041 \\
\midrule
\multirow{4}{*}{GMM}	& Asymp	& C	& & .044 & \multirow{4}{*}{.056} & .068 & .024 \\
									 & Asymp		& MR &	& .021 &  & .091 & .070 \\
			  						 & Boot (sym) & MR & L & .003 &  & .108 & .105 \\
			  						 & Boot (eqt) & MR & L & .019 &  & .115 & .096 \\
\midrule
\multirow{4}{*}{EL}			 & Asymp 	 & C 	& & .058 & \multirow{4}{*}{.068} & .079 & .021 \\
			  						 & Asymp 	  & MR	& & .056 &  & .080 & .024 \\
			  						 & Boot (sym) & MR & L  & .041 & & .096 & .055 \\
			  						 & Boot (eqt) & MR & L  & .049 & & .099 & .050 \\
\midrule
\multirow{4}{*}{ET}			 & Asymp 	 & C  &	& .058 & \multirow{4}{*}{.070} & .081 & .023 \\
			  						 & Asymp 	  & MR	& & .052 &  & .087 & .035 \\
			  						 & Boot (sym) & MR & L & .035 & & .105 & .070 \\
			  						 & Boots (eqt) & MR & L & .048 &  & .110 & .062 \\
\midrule
\multirow{4}{*}{ETEL}		& Asymp 	 & C  &	& .060 & \multirow{4}{*}{.071} & .083 & .023 \\
			  						 & Asymp 	  & MR	& & .056 &  & .086 & .030 \\
			  						 & Boot (sym) & MR & L &  .039 &  & .104 &  .066\\
			  						 & Boot (eqt) & MR & L & .051 &  & .108 & .057 \\			  
\bottomrule
\end{tabular}
\caption{95\% Confidence Intervals for the Returns to Schooling. Number of Bootstrap Repetition $B=5,000$.}
\label{tb_CI}
\end{table}

\clearpage
\thispagestyle{empty}

\end{document}